*Review Article*

# Machine and Deep Learning for IoT Security and Privacy: Applications, Challenges, and Future Directions

Subrato Bharati[1], Prajoy Podder[2]

[1,2]Institute of Information and Communication Technology (IICT), Bangladesh University of Engineering and Technology (BUET), Dhaka-1205, Bangladesh

Correspondence should be addressed to Subrato Bharati; subratobharati1@gmail.com

**Abstract:** The integration of the Internet of Things (IoT) connects a number of intelligent devices with a minimum of human interference that can interact with one another. IoT is rapidly emerging in the areas of computer science. However, new security problems were posed by the cross-cutting design of the multidisciplinary elements and IoT systems involved in deploying such schemes. Ineffective is the implementation of security protocols, i.e., authentication, encryption, application security, and access network for IoT systems and their essential weaknesses in security. Current security approaches can also be improved to protect the IoT environment effectively. In recent years, deep learning (DL)/ machine learning (ML) has progressed significantly in various critical implementations. Therefore, DL/ML methods are essential to turn IoT systems protection from simply enabling safe contact between IoT systems to intelligence systems in security. This review aims to include an extensive analysis of ML systems and state-of-the-art developments in DL methods to improve enhanced IoT device protection methods. On the other hand, various new insights in machine and deep learning for IoT Securities illustrate how it could help future research. IoT protection risks relating to emerging or essential threats are identified, as well as future IoT device attacks and possible threats associated with each surface. We then carefully analyze DL and ML IoT protection approaches and present each approach's benefits, possibilities, and weaknesses. This review discusses a number of potential challenges and limitations. The future works, recommendations, and suggestions of DL/ML in IoT security are also included.

## 1. Introduction

Internet of Things (IoT) considers the interconnection between several devices, i.e., industrial systems, intelligent sensors, autonomous vehicles, mechanisms and terminals, mechanical systems, etc. [1, 2]. Alternatively, it can be termed as a network of physical things or objects that are connected with limited communication, computation, and storage capabilities along with embedded electronics (i.e., sensors and actuators), connectivity of network, and software that enables these things to exchange, analyze, as well as collect data [3]. IoT relates to our everyday life, extending from smart devices in the household, i.e., smart meters, IP cameras, smoke detectors, smart adapters, smart refrigerators, smart bulbs, AC, smart ovens, and temperature sensors, to more advanced devices, for example, heartbeat detectors, radio-frequency identification (RFID) devices, accelerometers, IoT in automobiles, sensors in rooms, etc. [4]. Several services and applications referred to by the IoT are emerging in personal healthcare, home appliances, critical agricultural infrastructure, and the military [1].

The massive scale of IoT networks introduces latest issues, including the management of these devices, the complete volume of data, communication, storage, processing, as well as security and privacy concerns, among others. There has been substantial research into the various components of the IoT, such as architecture, communication, applications, protocols, security, and privacy, to name a few. The guarantee of security and privacy and user satisfaction are the cornerstones of the commercialization of IoT technology. The fact that the IoT makes use of empowering technologies including cloud computing (CC), software-defined networking (SDN), and edge computing enhances the number of dangers that attackers can encounter. As a result, monitoring security in the development of IoT infrastructure has become challenging and complex. Solutions must consist of wide-ranging considerations to fulfill the security challenges [5]. On the other hand, IoT systems are frequently put to use in an unprepared state. As a result, a fraudster can use wireless networks to connect to IoT devices and gain physical access to confidential data. Complexity and integrative arrangements characterize IoT systems. In light of the proliferation of connected devices, it might be difficult to meet the ever-evolving security standards for the IoT. In order to provide the necessary level of security, solutions need to consider the system as a whole. However, most IoT devices can function independently of human input. Someone without permission may thus acquire physical access to these devices [6-8].

Furthermore, The IoT system introduces novel attack surfaces. The interconnected and interdependent systems cause these types of attacks to surface. Accordingly, the security of IoT systems is faced with a higher risk than the security of other traditional computing devices. The outdated computing systems will be fruitless for these IoT schemes [9-11].

IoT systems ought to instantaneously consider security, energy efficiency, IoT software applications, and data analytics at the time of related tasks as a sign of the wide-ranging application [7]. This expansion offers an innovative scope for scholars from the interdisciplinary research program to consider recent challenges in the IoT schemes from various perceptions. However, the large-scale as well as cross-cutting nature of IoT devices and the many components engaged in their implementation, have created new security issues. The IoT devices characteristic presents various security issues. Additionally, the stages of IoT provide a massive number of useful information. If this information is not analyzed and transmitted securely, a crucial privacy gap may occur. Applying related security mechanisms, such as authentication, encryption, application security, network security, and access control, is inadequate and challenging for enormous schemes with numerous associated schemes. Every portion of the IoT platform contains intrinsic vulnerabilities. Such as, a special kind of botnet like 'Mirai' has newly affected extensively distributed denial of service (DDoS) attacks by using IoT systems [9, 12].

For extensive methods with multiple connected devices and each module of the method having inherent vulnerabilities, it is difficult and inadequate to apply existing security protection mechanisms like encryption, identity verification, application security, access control, and computer security [9]. For example, the 'Mirai' botnet has lately been responsible for large-scale DDoS attacks by abusing IoT devices. For the IoT ecosystem, existing security methods need to be improved. However, the deployment of cryptographic functions against a particular security issue is rapidly overtaken by new categories of attack developed by the attackers in order to bypass current remedies. Addresses spoofed source IP are commonly applied in magnified DDoS attacks to hide the location of attack's from the targeted organization's security teams. As a result of the vulnerabilities in IoT systems, more sophisticated and catastrophic attacks such as Mirai might be predicted. On account of the wide range of IoT scenarios and applications, knowing which security solutions are best for IoT systems is not easy. As a result, the focus of the study should be on devising appropriate IoT security methods [12, 13].

While security and privacy are interconnected, security may exist without privacy, but privacy cannot exist without security. Security safeguards the availability of information, integrity, and confidentiality, while privacy is more detailed about privacy rights in relation to personal information. Regarding the processing of personal data, privacy takes precedence, but information security entails preventing illegal access to information assets. Personal data may relate to any information about a person, including names, credentials, addresses, social security numbers, bank account data, etc.

A number of ways have been suggested to address the boundary between security and privacy concerns in DL and ML. Homomorphic encryption, differential privacy, trusted execution and secure multi-party computing environment are the four most often used DL and ML privacy technologies. This technique uses differential privacy to prevent the adversary from figuring out which instances were utilized to build the target model. Training and testing data are protected by safe multi-party computing and homomorphic encryption. For sensitive data security and training code, trusted execution environments leverage hardware-based security and isolation. These approaches, on the other hand, greatly increase the computing burden and need a tailored approach for each type of neural network. DL or ML privacy concerns have yet to be addressed in a way that is accepted worldwide. To protect against adversarial attacks, a wide variety of security measures have been suggested, which may be divided into three categories: input pre-processing, strengthening the model's resilience and malware detection. Preprocessing's goal is to lessen the model's reliance on immunity by doing operations such picture transformation, randomization, and denoising that don't often need model update or retraining. Introducing regulation, feature denoising, and adversarial training as well as other techniques to strengthen the model's robustness via model retraining and change, fall under the second group. Adaptive denoising and image transformation detection are the examples of third-category detection mechanisms that may be implemented before the first layer of the model. To the best of our knowledge, no defense strategy exists that can entirely protect against adversarial cases despite the many defensive mechanisms that have been offered. To counter hostile instances, adversarial training is currently the most effective technique. For poisoning attacks, there are two basic means of defense. The first is an outlier identification technique, which eliminates outliers from the relevant set. The second step is to enhance the neural network's ability to withstand contamination from poisoned samples.

## 1.1. Motivation and Scopes

Deep learning (DL) and machine learning (ML) are effective methods of data analytics as well as investigation to realize 'abnormal' and 'normal' behavior following how IoT devices and components interrelate with each other within the environment of IoT [14]. The IoT systems' input data can be investigated and collected to find out standard patterns of the interface, thus detecting malicious manners at the initial stages. Furthermore, DL/ML techniques can be significant in identifying new threats, which are regular modifications of existing threats, as they can highly detect upcoming unknown threats by learning from previous attacks. As a result, IoT systems need to be able to move from secure communication between security-based intelligence and devices via ML/DL techniques for safe and efficient systems.

Several unique properties of IoT networks will be discussed in the following paragraphs.

**Heterogeneity:** Each item in an IoT network has unique features, communication protocols, and capabilities that all function together. Different communication paradigms and protocols (such as Ethernet or cellular), as well as varied hardware resource limits, might be used by the devices. On the one hand, this diversity allows devices to communicate across platforms, but on the other, it introduces additional obstacles to the network of IoT.

**Proximity communication:** Additionally, IoT devices may communicate with one other without trusting on a central authority like base stations, which is an important feature. Dedicated short range communication (DSRC) and other point-to-point communication technologies are used in Device-to-device communication (D2D). Decoupling services and networks allows device-centric and content-centric communication, broadening the IoT service spectrum, whereas the conventional Internet's design is more network-centric.

**Massive deployment:** Massive deployment is predicted that the existing internet's capabilities would be exceeded by the billions of devices linked to it and the Internet. Massive IoT deployments are not without their own set of difficulties. Storage and architecture networks for intelligent devices, efficient protocols for data transfer, and proactive detection and protection of IoT-based devices from malicious attacks are only some of the difficulties that need to be addressed. A worldwide information and communication infrastructure that can be retrieved from everywhere as well as at any time is envisaged for IoT devices. How much is connection reliant on the kind of IoT service and application provided? For example, a swarm of sensors or a connected automobile may have a local connection, whereas critical infrastructure management and smart home access through mobile infrastructure may have global connectivity.

**Low-cost and low-power communication:** For optimal network operations, low-cost as well as ultra-low-power solutions are needed for the massive networking of IoT devices. For modern and critical IoT connectivity, self-healing and self-organization features are needed. Self-organizing networks ought to be implemented in these cases since relying on the network structure is not an option.

**Low Latency and Ultra-Reliable Communication (LLURC):** Remote surgical procedures, intelligent transportation, and industrial process automation systems all rely on the ability of IoT networks to reliably and quickly respond to real-time demands.

**Safety:** As a result of the enormous number of IoT devices linked to the internet, the private data exchanged via these systems may be at risk. Privacy and device security are also vital considerations. One of the most exciting aspects of IoT is its ability to make timely and intelligent choices based on the data it processes.

**Dynamic changing network:** An enormous number of IoT devices need proper management. These devices may behave dynamically. For example, when a device goes to sleep or wakes up, it is determined by various factors, including the software it is running.

The commercialization of IoT services and applications is heavily dependent on security and privacy. Various sectors, such as health care and business, have been impacted by security breaches that range from basic hacking to well-coordinated intrusions at the corporate level. Due to their restrictions and the environment in which they operate, IoT devices and apps face significant security issues. IoT security and privacy concerns have been thoroughly considered from different viewpoints, including communication privacy and security, architecture data security, and identity management as well as malware analysis [4]. Sections 3 and 5 explore more into the issues of security and the threat model.

According to Fernandes et al. [15], security challenges in IoT and conventional information technology (IT) devices are comparable and different. In addition, they also addressed privacy concerns. Software, hardware, networks, and applications are some of the most often cited points of comparison and contrast in this debate. IoT and conventional IT have a lot of things in common when it comes to security concerns. Despite this, the real concern of the IoT is the lack of available resources, which makes it challenging to implement advanced security measures in IoT networks. Improved algorithms and cross-layer architecture are also needed to address IoT privacy and security concerns. As part of an overall privacy and security method for IoT, current security solutions will be nominated for consideration, as well as new intelligent, resilient, scalable, and evolutionary methods to handle IoT security concerns.

ML implies intelligent procedures that utilize previous experiences or example data to understand how to maximize performance criteria. Algorithms that use machine learning to develop behavioral models on massive datasets are known as ML. Because of machine learning, computers can learn independently, even if no instructions are provided. The newly included data is fed into these models, which serve as a foundation for generating predictions about the future. AI, optimization, information theory, and cognitive science all have origins in ML, so it is a multidisciplinary field of study [16]. ML is no exception. Robotics, voice recognition, and other areas where people are unable to apply their skills, such as hostile environments, need the use of machine learning [17]. It may also be used when the answer to a particular issue evolves over time. To put this into context, Google utilizes ML to identify risks to mobile devices and apps running on the Android platform. Infected mobile devices may be scanned and cleaned using this tool. Macie, an Amazon tool that applies ML to organize as well as categorize data stored in Amazon's cloud storage, was also released recently. False positives and true negatives may occur even when ML methods are used correctly. As a result, if an incorrect prediction is produced, ML approaches need direction and change of the model.

Contrary to popular belief, with DL, a new kind of ML, the model is able to establish its accuracy of prediction. Prediction and classification tasks in novel applications of IoT with customized and contextual support might benefit from the self-service character of DL models. Moreover, the complete volume of data produced by IoT networks necessitates the use of DL and ML methods to offer intelligence to the systems. In addition, the IoT data created by DL and ML algorithms can be effectively exploited to make educated and intelligent choices by the IoT systems. The analyses of privacy, security, malware, and attack detection are just a few of the many applications for DL and ML. DL methods can also be employed in IoT systems to conduct identification tasks and complicated sensing to develop new apps and services that take into account real-time interactions among people, the physical environment, and smart devices.

Real-world uses of ML in security include the following:

(i) Different handwriting styles are used for character recognition in security encryption.

(ii) Recognition of faces in forensics: lighting, pose, occlusion (beard, glasses), hairstyle, make-up, etc.

(iii) Software and apps that contain malicious code need to be identified.

(iv) Behavior analysis is used to identify DDoS attacks on infrastructure. On the other hand, there are several difficulties associated with applying DL and ML in IoT applications. For example, designing an appropriate model for processing data from many IoT applications is challenging. In the same way, correctly classifying input data is likewise a complex undertaking. The use of little marked data in the learning process is also tricky. Using these models on IoT devices with limited processing and storage resources presents further difficulties [18]. Like essential infrastructure and real-time applications, DL and ML algorithms produce anomalies. IoT security solutions that use DL and ML must be thoroughly analyzed in this context.

### 1.2. Contributions

The main influences of this work are presented below:

- A review of various types of attacks with its example is discussed.
- Comprehensive analysis of ML and latest developments in IoT defense DL methods: the most promising DL and ML algorithms are examined for IoT protection schemes, and their benefits, drawbacks, and implementations are addressed in the security of IoT systems. In addition, compare, and description tables are provided for DL and ML approaches for learning lessons.
- A number of state-of-art applications of DL and ML in IoT security and privacy are illustrated.
- We offer a taxonomy of the most recent IoT privacy and security solutions based on deep learning and machine learning techniques. Moreover, new insights of ML and DL in IoT securities are illustrated.
- Potential limitations, challenges, future directions, and suggestions of DL and ML are appeared, how they could help the recent and future research.

The work is presented as follows: Literature reviews with their limitations and why this work is needed are illustrated in Section 2. Next, applications of ML or DL to IoT threats are illustrated in Section 3. Moreover, Section 4 appears in the DL and ML models, where we can find how to work with each ML and DL model in IoT security, and solutions are also described in Section 5. In section 6, we can see a number of new insights into Deep and Machine Learning for IoT Security that can help future research works. Section 7 discusses challenges, limitations, and future directions. A number of suggestions and recommendations are presented in Section 8, and Section 9 shows the conclusion.

### 2. Literature Reviews

A number of surveys or reviews have covered IoT security to offer some guidelines for future challenges. Though several studies have looked at IoT security, none have focused on DL or ML applications for IoT security. Several works [19-25] have been reviewed for motivating and organizing the challenges in access control, authentication, application security, encryption, and network security in IoT environments. The survey of [26] provided a survey of IoT communication on security issues with its solutions. Another paper [27] emphasized IoT systems for intrusion detection.

Moreover, IoT frameworks for regulatory approaches and legal issues can determine security and privacy requirements [28]. The context of distributed IoT has also covered privacy and security in [29]. These works also concerned various challenges. Several issues must be found out, and the researchers assert that the distributed IoT method offers numerous advantages in terms of privacy and security. The survey of [30] described evolving threats and vulnerabilities in IoT devices, for example, threats of ransomware as well as security concerns. The authors of [31] concisely indicated the context of IoT using ML techniques concerning data security and privacy protection. This survey also described three challenges with respect to ML application in IoT environments (i.e., communication and computation overhead, partial state consideration, and backup security justifications). Numerous survey studies, including [31, 32], have examined the use of data mining and ML methods in cybersecurity to assist intrusion detection. Above all, they reviewed anomaly detections and misuse in cyberspace [32]. The methodology was based on several classes of AI (artificial intelligence) methods from the point of view of an IoT context, and the opportunities of applying those approaches in IoT environments were observed. However, that work did not emphasize the implementation of DL an IoT perceptive. The review of [3] also provided ML techniques in IoT security where they offered future challenges and current solutions. Another survey of ML methods for wireless sensor networks (WSNs) was appeared in [33].

The motivation of that study was to review ML methods in real-life WSN applications. i.e., clustering, localization, routing and unrealistic aspects of quality of service (QoS) and security. The framework of WSNs in DL methods was described in the work of [34]. On the other hand, this work emphasizes network configuration. Besides, it differs from the proposed survey that focuses on DL/ML methods for ensuring IoT security. Some traditional ML techniques [35] were considered with advanced methods, including DL methods [36] for processing big data. Above all, the relationship of several ML techniques for signal processing approaches was focused on investigating and processing relevant big data. An overview of DL is offered on state-of-the-art approaches [37]. The survey proposed the opportunities and challenges of various existing solutions with their uses and evolution. The essential principles of several DL classifiers were evaluated with their procedures in addition to developments of DL methods in several uses [38, 39], for example, speech processing, pattern recognition, and computer vision. In mobile advertising, a review of improvement in DL methods was used for recommendation systems, which show a crucial role [40]. Various effective ML applications [41] were similarly conducted in self-organizing networks. The survey focused on the merits and demerits of various methods and offered future opportunities and challenges in expanding artificial intelligence and future network design [42]. The significance of 5G in artificial intelligence was highlighted. Intrusion detection using data mining was covered in [43]. Application of multimedia mobile was surveyed conducting DL methods as well [44]. Recent DL techniques in mobile security, speech recognition, mobile healthcare, language translation, and ambient intelligence were focused. Similar research was conducted on the most advanced, state-of-the-art deep learning approaches used in a diversity of IoT data analytics applications [45]. On the other hand, our survey covers a complete review of recent progress in deep learning approaches and cutting-edge machine learning approaches for ensuring security in IoT. This review compares and identifies the advantages, prospects and weaknesses of different DL/ML approaches for security in IoT. This paper also discusses numerous future directions and challenges and discuss the realized problems and future prospects based on a study of possible DL/ML applications in the context of IoT security, thus offering an effective guideline for researchers or scholars to modify the security of IoT environment from simply empowering a secure communication between IoT modules to end on IoT security on the basis of intelligence methods.

3. IoT Threat

Several heterogeneous sensing systems communicating with one another through a local area network (LAN) are referred to as the IoT [46]. The risks in IoT are distinct from those posed by traditional networks, owing in large part to the capabilities accessible to end devices [47]. The traditional Internet relies on powerful computers and servers with plenty of resources, while the IoT relies on equipment with low memory and computing power. That is why an IoT device in the real world cannot continue employing multifactor authentication and dynamic protocols like a regular network. Wireless protocols applied by IoT devices, for example, ZigBee and LoRa are less secure than those used by traditional networks. A lack of a standard operating system and particular features inside IoT applications have resulted in various data contents and formats in the systems, making the creation of a uniform security protocol complicated [48]. There are several security and privacy problems associated with the IoT because of these flaws. As a network grows in size, the risk of an attack rises. Since the IoT has no firewalls, its network is more vulnerable than a traditional office or company network. IoT systems that exchange data with one another are frequently multi-vendor systems, adhering to a wide range of spectra and protocols from different manufacturers. The connection between such devices is difficult, necessitating the use of a trustworthy third party as a bridge [49]. Additionally, many reports have posed concerns about how billions of smart devices receive app updates [50, 51]. Since an IoT device has small computing resources, its ability to cope with advanced threats is harmed. To conclude, IoT weaknesses may be classified as either essential or widespread. For example, although vulnerabilities such as battery drain attacks, insufficient standardization, and insufficient confidence are exclusive to IoT systems, vulnerabilities in internet-inherited systems may be considered general. Numerous IoT risks have been identified and classified in the past [32, 52-55]. We address the most often identified challenges to the IoT over the last ten years and try to categorize them into privacy and protection classifications. Privacy and security are basic principles that turn around network availability [56-58]. On the internet of things, data may take several forms, including a user's identification records, an order issued to a car by a keyfob, or a graphical chat between two people. Unauthorized data disclosure can constitute a breach of data security, integrity, or availability. If a threat compromises secrecy, it is classified as a privacy threat. Both data confidentiality and network stability are jeopardized by security attacks. Fig.1 depicts the various types of threat that exist in IoT domains.

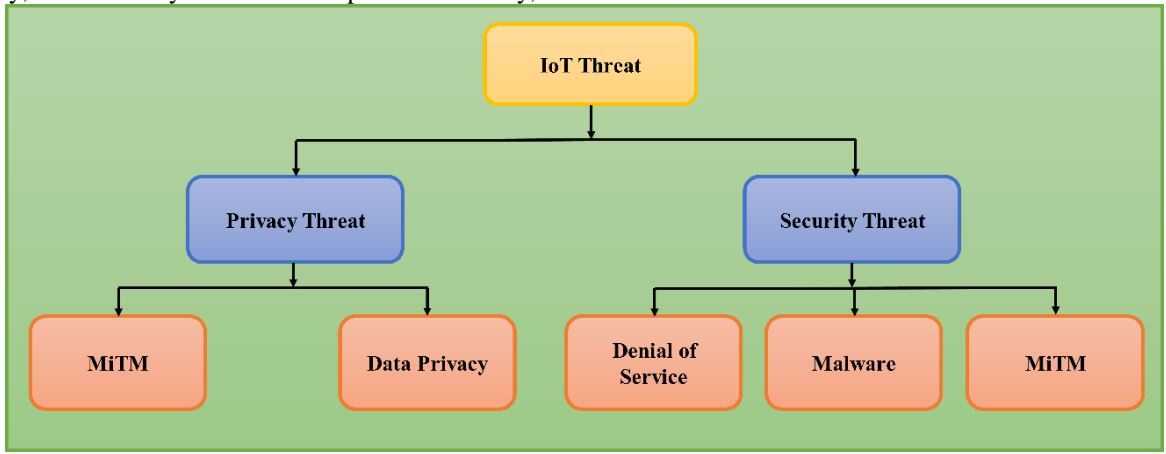

**Figure 1:** Sorts of IoT threat

### 3.1. Privacy Threats

Along with protection risks, common privacy threats against IoT data and their users, i.e., de-anonymization, inference, and sniffing. In any scenario, the effect is on data secrecy, regardless of whether the data is at rest or in motion. This segment discusses different types of privacy threats.

#### 3.1.1. Man-in-the-middle (MiTM)

Passive MiTM Attacks (PMA) and Active MiTM Attacks (AMA) are two types of MiTM attacks. The PMA passively monitors data flow between two systems. The data is not altered if the PMA violates anonymity. An intruder with access to a computer will observe passively for months before launching an attack. With the proliferation of cameras in IoT devices such as dolls, wristwatches, and tablets, the impact of Passive MiTM Attacks such as sniffing and eavesdropping is immense. In comparison, the AMA is actively engaged in data misuse, dealing with an operator pretending to be someone else, e.g., improperly, impersonation to retrieve a profile, or an authorization attack.

#### 3.1.2. Privacy in data

As with MiTM attacks, attacks in data privacy are categorized as Passive Data Privacy Attacks (PDPA) or Active Data Privacy Attacks (ADPA). Data tampering, data outflow [59], re-identification, and identity stealing are all issues relating to data protection [60]. Re-identification attacks often referred to as hypothesis attacks, are focused on position recognition, de-anonymization, and data aggregation [60]. The primary objective of these attacks is to collect data from various outlets to discover the targets' characters. Certain attackers can conduct the data gathered to mimic a specific goal [54]. Each attack that modifies records, such as data tampering, falls under the definition of ADPA, while data leakage and re-identification fall under the definition of PDPA.

### 3.2. Security Threats

#### 3.2.1. Denial of Service

While compared to other types of security threats, denial-of-service (DoS) has the simplest application. In addition, as several IoT devices with poor security features continue to increase, DoS attacks are an attacker's favorite tool. The primary goal of a DoS attack is to overwhelm the IoT network with illegal requirements and to deplete network resources, including bandwidth. As a result, legal consumers cannot access the services. DDoS is a more complex form of DoS attack where a particular objective is attacked from several origins, which makes the attack more difficult to detect and avoid [61-66]. Though DDoS attacks have different flavors, they all have a similar objective. A variety of attacks in DDoS include SYN floods [67], in which a hacker dispatches a number of SYN appeals to a remarkable target; attacks in internet control message protocol (ICMP) [68] (in which several ICMP packets are being transmitted via a spoof-IP); crossfire attacks [69] in which an attacker is attacking a complex, massive botnet; and User data logs (User Datagram Protocol). Botnet attacks [70] are a form of DDoS attack occurring in IoT networks. A botnet is a group of IoT devices hacked to start an attack on a particular item, such as a bank server. Botnet attacks can be carried out by various protocols, including Message Queuing Telemetry

Transportation (MQTT), a Domain Name Server (DNS), and a Hypertext Transfer Protocol (HTP), as outlined in [70]. Some ways to detect DoS attacks in an IoT environment are proposed. The authors of [62] showed how an attack in a fog-to-things environment is identified by applying DL techniques. In a second paper, the work of [61] suggested that the usage of distributed DL on fog computing could ease DDoS attacks. The IDS is a sequence of development exercises to lessen attacks in DDoS with sophisticated computer learning and deep learning algorithms [64, 65]. The Software Defined Networks flood issue has been emphasized by the authors of [63] and [66]. The study showed that the top layer of the SDN is susceptible to a brute force attack because of the insufficient protection in the TCP channel of plain text.

### 3.2.2. Malware

One of the most well-known attack domains is the execution and injection of malicious code into IoT systems via developing existing vulnerabilities in IoT systems. Vulnerabilities in application security, authentication, and authorization may be exploited for malware injection. Without these approaches, physically tampering with IoT devices to modify the software and misconfiguring security parameters may also enable attackers to introduce malicious code. Malware is a persistent threat that is executed via various methods due to the vulnerabilities mentioned above. Malware comes in a variety of forms, including spyware, bot, adware, ransomware, virus, and trojan, to mention a few [48, 71]. Moreover, Azmoodeh et al. [72] conducted research on malware distributed through the Internet of Battlefield Thing (IoBT). These hackers are often well-trained, well-funded, and state-sponsored. The authors of [73], [74], and [75] used various supervised machine learning algorithms to attempt to protect resource-constrained android devices against malware attacks. The studies [51, 76, 77] examined malware detection in-depth and identified many security flaws in the Android framework, specifically at the application layer. It contains applications with a variety of component forms.

### 3.2.3. Man-in-the-middle

Man-in-the-middle (MiTM) attacks were among the first cyber threat types [78]. Spoofing and impersonation include MiTM attacks. For example, the MiTM attacker could communicate with a node "X", which communicates with destination "A". Similarly, a hacker can use this kind of attack to link to a server with an HTTPS connection in SSL stripping. Recently, several researchers have been motivated to develop security against MiTM attacks [79-82]. A work of [79] illustrated the clinical condition in which a patient is given an insulin injection instantly. This form of program is subject to a fatal MITM attack. Likewise, the authors of [81] addressed new safety methods for wireless mobile devices, including the usage of a concealed key, in the face of impersonation attacks. Using non-volatile memory and cryptography using hash values, this key has been safeguarded against loss or theft. This method was not only insecure, but it was also wasteful of resources, like OAuth 2.0, which is the most extensively used IoT protocol and is vulnerable to cross-site request forgery attacks (CSRF). The OAuth protocol takes a considerable time to authenticate computers physically. The researchers of [82] listed a physical layer security defect in the authentication of a wireless system. They discussed the current hypothesis test, which compares some information in radio channels to the channel record of alice to identify Eve spotter in wireless networks. It is sometimes inaccessible, mainly in active networks.

### 3.3. Another threat to privacy and security

There are two types of security threats: physical and cyber. Active and passive cyber risks are further sub-classified. The following part provides an outline of some of these risks and threats.

### 3.3.1. Physical Threats

Physical destruction is one kind of threat that may be posed. A cyber-attack is not usually possible in these cases since the attacker lacks the necessary technical know-how. To put it another way, the attacker can only impact the IoT devices that can be physically accessed by the hacker. If IoT systems are used widely, these sorts of attacks may become more widespread since cameras and sensors are projected to be widely available and accessible. [29, 83]. Natural catastrophes, such as earthquakes or floods, or human-caused disasters, such as wars, may also create physical threats [84, 85].

### 3.3.2. Cyber Threats

**Active threats:** As part of an active threat, the attacker is not only skilled at listening in on communication channels but also at changing IoT devices to change settings and regulate communication, refuse services, and many more. A series of interventions, interruptions, and alterations may be used in an attack. There are several ways to attack an IoT system, including impersonation (such as spoofing), data tampering, malicious inputs, and DoS. IoT devices or authorized users may be impersonated in a cyberattack called an impersonation attack. If an attack vector is available, active intruders may try to mimic an IoT entity in part or its whole. An IoT system is attacked using a malicious input attack in order to introduce malicious software into the system. Code injection attacks will be carried out using this program. Malicious software injected into IoT systems has a dynamic character, and new attack types are continually being produced to breach the systems' IoT components in remarkable ways [30, 86]. Data tampering, on the other hand, is the act of purposely altering (deleting, altering, modifying, or manipulating) data via illegal actions. Transmitting and storing data are commonplace. IoT systems may be compromised in both cases, which might have substantial consequences, i.e., altering the IoT-based billing price of the smart grid. IoT may be subjected to a wide range of denial-of-service attacks. A wide variety of DoS attacks may be found, from those that target Internet traffic to those that target cellular connectivity. It is more difficult to distinguish a DDoS attack from regular traffic and devices than a DoS attack with a small number of devices or a large signal, which is more straightforward to distinguish from regular devices and traffic than DDoS attacks. A typical goal of DoS attacks is to disrupt the availability of IoT services [27]. Many IoT systems are vulnerable to devastating DDoS attacks, such as Mirai, since they include billions of linked devices. Using IoT devices, the Mirai botnet has lately been utilized to launch large-scale DDoS attacks.

**Passive threats:** Eavesdropping on the communications network or the channels are all that is required to carry out a passive threat. An eavesdropper may obtain sensor data, monitor the sensor bearers, or both by listening in on their conversations. The illegal market for important personal information, such as health data, has exploded recently [87]. In comparison to credit card information, which sells for $1.50, and social security numbers, which sell for $3, personal health information is worth $50 on the black market. Furthermore, an attacker may determine the location of an IoT device's owner by eavesdropping on the owner's communications if they are in range [88, 89].

**4. Applications of IoT security**

Almost all IoT applications, whether currently in use or under development, have security as a top priority. IoT applications are growing at a fast pace and have already penetrated most of the current sectors. A few IoT applications required stricter support of security from IoT-based technologies they employ, despite operators supporting these apps with current networking technology. There are several crucial IoT applications that are covered in this section.

**4.1. Home Automation**

IoT has a broad range of applications, including home automation. This category includes applications for remotely controlling electrical appliances to save energy, devices put on doors and windows to find intruders, and more. Energy and water use are being tracked via the use of monitoring devices, and customers are being counseled on ways to conserve resources and money. The authors of [318] have recommended the application of security techniques that is based on logic to improve home security. Users' activities in crucial areas of the house are being compared to their typical behavior in order to identify intrusions. Attackers may, however, get access to IoT devices in the house without the owner's permission and use that access to do damage to the owner. For example, the number of burglaries has skyrocketed after different home automation systems were installed [318]. Internet traffic to and from a smart home has been used by opponents before to determine a person's activities or even their presence at the residence.

**4.2. Smart Cities**

Smart cities make full use of newly available computing and communication technologies in order to develop the lives of its residents [319]. Smart cities, smart transportation, smart disaster response, and other smart services are all included. Governments throughout the globe are promoting the creation of smart cities via different incentives [320]. Even while smart apps are meant to enhance the quality of life for individuals, they pose a risk to their privacy. Citizens' credit card purchasing habits and information might be at danger while using smart card services. Smart mobility apps can expose where its users are located. Parents can keep an eye on their children with the use of mobile apps. The child's security may be jeopardized, though, if these applications were hacked.

**4.3. Smart Retail**

Applications of IoT are widely employed in retail. Many applications have been developed to track the temperature and humidity of inventory as it moves through the supply chain. Additionally, IoT may be utilized to optimize warehouse refilling by monitoring goods movement. Intelligent shopping apps are also being generated to help consumers based on preferences, habits, and sensitivities to certain components, for example. These applications are being developed. An augmented-reality system that allows physical shops to experience internet buying has also been created. IoT applications deployed and used by retail firms have been plagued by security concerns. They include Home Depot, Apple, Sony, and JPMorgan Chase. Attackers may attempt to breach IoT apps related to the storage conditions of goods and communicate incorrect information about the items to customers in an effort to promote sales. Consumers' credit and debit card information, email addresses, phone numbers, and other personal information can be stolen if the elements of security are not comprised of smart retail. This can result in monetary losses for both the businesses and the customers.

**4.4. Animal Farming and Smart Agriculture**

Soil moisture monitoring, maintaining selective watering in dry zones, micro-climate conditions, and controlling temperature and humidity are only a few of the smart farming practices. The use of sophisticated features in agriculture may lead to higher yields and assist farmers avoid monetary losses. Moreover, fungus and other microbiological pollutants may be prevented by carefully monitoring and controlling the humidity and temperature levels in different vegetable and grain production processes. The quality and quantity of vegetables and crops may also be improved by controlling the climate. As with crop monitoring, apps in IoT are available to track the activity and health status of farm animals using sensors attached to the animals. Compromised agricultural apps might lead to animal theft and crop harm.

### 4.5. Smart Grids and Smart Metering

Management, monitoring, and measurements may all be done with smart meters. It is the most prevalent use of smart meters in smart grids, where power use is tracked and measured. A smart metering system might potentially be used to combat power theft. Storage tank monitoring and cistern level monitoring are two further uses for smart meters. By dynamically adjusting the position of solar panels in the sky, smart meters may also be used to monitor and enhance the performance of solar energy facilities. Other applications of IoT include smart meters to monitor water pressure to measure the weight of items or in water transportation systems. Smart meters, nevertheless, are susceptible to both cyberattacks and physical attacks compared to traditional meters, which can only be interfered with by physical means. Smart meters or advanced metering infrastructure (AMI) are designed to carry out additional tasks beyond the simple tracking of energy use. A smart home area network (HAN) connects all of a household's electrical appliances to smart meters, which may be used to monitor use and costs. Consumer or adversary intrusions into such systems may alter the acquired data, resulting in financial losses for users or service providers [321].

### 4.6. Smart Environment

IoT may be applied to identify forest fires, monitor snow levels at high altitudes, prevent landslides, detect earthquakes early, monitor pollution, and many other things. There is a strong connection between the lives of humans and animals in these regions and the use of IoT applications. The information from these applications of IoT will also be used by government entities working in these domains. The repercussions of a security breach or vulnerability in any IoT application area might be dire. False negatives and false positives may have severe effects for IoT applications in this situation. For example, if the app begins incorrectly identifying earthquakes, the government and companies may suffer financial damages. If, on the other side, the software fails to forecast the earthquake, both property and lives will be lost. As a result, security flaws and data manipulation must be avoided in smart environment applications.

### 4.7. Security and Emergencies

The deployment of numerous IoT applications in the field of security and emergencies is another key development. It covers applications such as restricting access to restricted areas to only those with proper credentials. Hazardous gas leak detection in industrial regions and near chemical companies is another use for this technology. There are a variety of buildings where sensitive information is stored on computers or where sensitive commodities are stored. Protecting sensitive information and items is possible with the use of security apps. Buildings with high levels of sensitivity, such as nuclear power plants, may benefit from the usage of IoT apps that monitor liquids. The repercussions of a security compromise in these apps might be dire. Criminals may, for example, attempt to get access to restricted regions by exploiting programs' security flaws. The immediate and long-term consequences of erroneous radiation level warnings may also be severe. Long-term radiation exposure in babies, for example, may cause significant disorders that are life threatening.

## 5. IoT applications of security threat for each layer

This section describes the four levels of an IoT application: the first is the sensing layer, the second is the network layer, the third is the middleware layer, and the fourth is the application layer. In an IoT application, each layer employs a variety of technologies that introduce a variety of challenges and security ricks with them.

Security risks in IoT applications are discussed in this section for the four tiers. This section also discusses the particular security concerns of the gateways that link these levels.

### 5.1. Sensing layer and its security issues

This layer focuses on physical actuators and sensors for the IoT. Sensors pick up on the physical activity taking on around them [322]. While sensors collect information about their surroundings, actuators take action on the physical world depending on that information. There are a variety of sensors that may be used to gather information, including video sensors, ultrasonic sensors, humidity and temperature sensors, and more. Chemical, electrical, electronic, mechanical sensors may all be employed to gather data about the physical world around us. IoT applications employ a variety of sensing layer technologies, including GPS, RFID, RSNs, WSNs, etc. Sensing layer security risks include the following:

(i) Injection Attack using Malicious Code: A malicious code is injected into the node's memory by the attacker. The firmware or software of IoT nodes is often changed over the air, which provides an entry point for malicious malware injection by attackers. Using malicious code, attackers may cause the nodes to execute certain attempts to get access or even undesired operations to the whole IoT system.

(ii) Node Capturing: Actuators and sensors are only two examples of low-power nodes in IoT applications. The opponents may launch a wide range of attacks against these nodes. A malicious node might be used to replace or capture the legitimate node in the IoT system. In reality, the attacker has complete control over the new node. This might lead to the full IoT application being compromised [323].

(iii) Side-Channel Attacks (SCA): There are a variety of side-channel attacks that may expose sensitive data besides those that target nodes directly. An adversary may get access to sensitive information via the microarchitectures of CPUs, electromagnetic emission, and their power consumption. Power consumption, laser-based, timing, and electromagnetic side channel attacks may all be used to launch attacks. To avoid side-channel attacks when implementing cryptography modules in modern circuits, there are many countermeasures.

(iv) Booting Attacks: Devices on the edge are susceptible to a variety of threats during starting up. A lack of built-in safety measures explains why this happens. When the node devices are rebooted, attackers may use this vulnerability to attack them. Due to their low power consumption and sleep-wake cycles, edge devices need to have a secure startup mechanism.

(v) Sleep Deprivation Attacks: Attackers' goal in these sorts of attacks is to drain the power supply of relatively powerless IoT edge devices. Due to a dead battery, the IoT nodes are unable to provide any service. For this to happen, malicious malware or increased power consumption on the edge devices is used to execute endless loops in those devices' CPUs and RAMs.

(vi) False Data Injection Attack: An attacker may then utilize the node to feed false data into the IoT system after it has been taken over. Inaccurate findings might lead to the IoT application malfunctioning. DDoS attacks may be launched using this strategy as well.

(vii) Eavesdropping and Interference: Open environments are widely used to launch IoT apps. This means that IoT applications are vulnerable to eavesdroppers as a consequence. It is possible for the attackers to eavesdrop and steal data throughout various stages of authentication or transmission.

### 5.2. Network layer and its security issues

The primary role of the network layer is to transfer data from the sensor layer to the computing unit for processing. The following are the most common network security problems.

(i) DoS/DDoS Attack: In this kind of attack, unwanted requests are issued in mass quantities to the attacked servers. As a result, legitimate users will be unable to access the targeted server's resources. DDoS attack occurs when several sources are exploited by the attacker to overwhelm the target server and cause it to become unusable. As a result of the complexity and variety of IoT networks, such attacks are more likely to take place because of this. Due to improper configuration, many IoT devices used in IoT applications are vulnerable to DDoS attacks. This vulnerability was used by the Mirai botnet attack, to block multiple services by sending requests to IoT devices that were poorly configured [12].

(ii) Routing Attacks: Nodes in an application of IoT that are attempting to conduct malicious activity may attempt to reroute traffic as it passes through the system. In a sinkhole attack, an adversary broadcasts a fake shortest route and actively encourages nodes to use it. Worm-hole attacks, when coupled with sinkhole attacks, may pose a major danger to computer security. Fast packet transmission may be achieved using a warm-hole connection between two nodes. An attacker may exploit a vulnerability in an IoT application by creating a "warm-hole" between a hacked node and a device on the internet.

(iii) Access Attack: The term "advanced persistent threat (APT)" may also apply to an access attack. In this form of attack, an unauthorized individual or an adversary obtains access to the IoT network without permission. For a long time, the attacker may remain unnoticed in the network. This kind of attack is aimed at stealing important information or data, rather than causing harm to the network. Applications of IoT are particularly vulnerable to attacks because they constantly collect and send crucial data.

(iv) Data Transit Attacks: The storage and sharing of data is a major concern for IoT applications. Because data is so valuable, it is constantly a target for cybercriminals and other bad guys. Even whether data is kept locally or in the cloud, it is subject to cyber attacks while it is in transit or is traveling between locations. Data travels a long way in IoT applications between actuators, sensors, and the cloud, among other places. Data transmissions using the IoT may be compromised since a variety of connecting mechanisms are in use.

(v) Phishing Site Attack: Phishing attacks are those in which a single attacker may target a large number of IoT devices with little to no effort. The attackers believe that at least some of the devices will succumb to the onslaught. There is a chance that individuals may encounter phishing sites when browsing the internet. Any IoT device that a user has access to becomes a target for cyberattacks as soon as their login credentials are stolen. Phishing attempts on the network layer of IoT are quite common [324].

### 5.3. Middleware layer and its security issues

The middleware in IoT is responsible for creating an abstraction layer between the application and network layers. It is also possible for middleware to offer substantial compute and storage capabilities [325]. The APIs provided by this layer are used to meet the needs of the application layer. The middleware layer contains machine learning, permanent data storage, brokers, queuing systems, and so forth. Middleware is important for providing a strong and dependable application of IoT, but it is also vulnerable to a variety of attacks. As a result of these attacks, the whole IoT application may be hijacked. Besides database and cloud security, middleware security is a major concern. These attacks on the middleware layer are described in more detail below.

(i) SQL Injection Attack: Middleware may be attacked using SQL Injection (SQLi), which is another attack vector. A malicious SQL query may be inserted into a program by an attacker in such attacks [326, 327]. This allows the attackers to access the private information of any user, as well as to change database entries themselves [328]. According to the Open Online Application Security Project (OWASP) top 10 2018 paper, SQLi is a top threat to web security.

(ii) Flooding Attack in Cloud: Cloud-based denial-of-service attacks use a similar methodology and impact QoS in the same way. The attackers use a continual stream of queries to a service to deplete cloud resources. By increasing the strain on the cloud servers, these attacks may have an important effect on cloud systems.

(iii) Man-in-the-Middle Attack: Subscribers and clients communicate with each other through the MQTT broker, which operates as a proxy for the MQTT protocol. Messages may be transmitted to several recipients without knowing where they're going thanks to this method's ability to disconnect the publishing server from the subscribers. As long as the attacker can take control of the broker, he or she will be able to take over all communication without the awareness of the clients.

(iv) Signature Wrapping Attack: XML signatures are utilized in the middleware's web services. An attacker may use weaknesses in Simple Object Access Protocol (SOAP) to break the signature scheme and perform operations or change intercepted messages in a signature wrapping attack.

**5.4. Gateways and its security**

Connecting devices, objects, people, and cloud services is a key function of a gateway. In addition, gateways aid in the provision of IoT device hardware and software. The decryption and encryption of IoT data, as well as the translation of protocols across various levels, are handled by gateways. ZigBee, LoraWan, TCP/IP and Z-Wave stacks are among the several IoT systems that are currently in use today. In the following, we'll look at some of the security issues that IoT gateways are facing.

(i) End-to-End Encryption: End-to-end application layer protection must be executed in order to guarantee data confidentiality [38]. Only the intended receiver may decode encrypted communications using this program. Zwave and Zigbee protocols offer encryption, however the gateways are necessary to decrypt and re-encrypt the messages in order to convert the information from one protocol to another. At the gateway level, this decryption exposes the data to security vulnerabilities.

(ii) Firmware updates: In order to obtain and install firmware upgrades, most IoT devices lack a user interface or the computational capacity. Typically, gateways are used to obtain and apply firmware upgrades. Verify signature validity and the existing and new firmware versions before implementing any changes.

(iii) Extra Interfaces: Installing IoT devices while keeping an eye on the attack surface is a critical technique [329]. IoT gateway manufacturers should only implement the protocols and interfaces that are strictly essential. A backdoor authentication or data leak should be prevented by restricting certain services and functionality for end users only.

(iv) Secure On-boarding: Whenever a new IoT sensor or device is added, encryption keys must be protected. All keys flow via gateways, which operate as a go-between for management services and new devices. During the on-boarding process, the gateways are vulnerable to eavesdropping attempts and MiMA aims at stealing the encryption keys.

**5.5. Application layer and its security**

The application layer is responsible for interacting with and serving customers directly. Smart cities, smart homes, smart grids, and other IoT applications all fall under this umbrella. This layer contains unique security concerns, such as data theft and privacy concerns, that are not present in other levels. Various apps have different security concerns at this tier. A sublayer between the network layer and application layer, known as a middleware layer or application support layer, is used in many IoT systems. There is a layer of support that enables different business services and aids in the allocation and calculation of resources. The application layer's most pressing security concerns are outlined here.

(i) Reprogram Attacks: If the process of programming the IoT devices is not secure, the devices might be reprogrammed remotely. This might eventually lead to complete control of the IoT [330].

(ii) Malicious Code Injection Attacks: To get access to a network or system, attackers often use the quickest or most straightforward technique. If the system is susceptible to misdirection and malicious scripts owing to poor code checks, then an attacker would select it as the first access point. XSS (cross-site scripting) is often used by attackers to introduce malicious code into a supposedly trustworthy website that is otherwise safe. For example, if an IoT account is hacked, it might cause the whole system to be rendered inoperable.

(iii) Access Control Attacks: The term "access control" is used to describe the practice of restricting access to a resource (such as an account or data) to just those who are allowed to use it. When a user's credentials are stolen, the whole IoT program is vulnerable.

(iv) Service Interruption Attacks: In the current literature, these attacks are also known as DDoS attacks or unlawful interruption attacks. IoT applications have been the target of a number of similar attacks in the past. By intentionally overloading the network or servers, these attacks prevent genuine users from accessing IoT apps.

(v) Data Thefts: Critical and confidential data is handled by IoT apps. In applications of IoT, there is a great deal of data mobility that makes it even more susceptible to attacks than data at rest. Unless the IoT apps are secure, consumers will be unwilling to provide their personal information. Data encryption, data isolation, privacy network, and management and user authentication and other methods and protocols are being employed to protect the applications of IoT against data theft.

## 6. Overview of DL and ML for IoT

Security and privacy are interdependent. It is possible to conceive of a setting that is safe yet does not provide individual confidentiality. One may envisage a dwelling that is private due to the presence of windows, yet this would not always provide protection from intruders. While privacy is impossible to get without sacrificing some level of security, the opposite is not true. Privacy is always compromised when security is inadequate or exposed. In this section, we will illustrate the most prominent ML and DL models for classifying security aspects in IoT. ML techniques refer to unsupervised and supervised methods. The supervised methods are classified into Naïve Bayes (NB), support vector machine (SVM), random forest (RF), k-nearest neighbors (KNN), decision tree (DT), ensemble learning (EL) and association rule (AR). Additionally, the unsupervised approaches only refer to two approaches including principal component analysis (PCA) and K-means. DL techniques are similarly classified into unsupervised, supervised, and hybrid methods.

### 6.1. ML in IoT Security

This section discusses the traditional ML algorithms. Fig. 2 depicts various ML classifiers for IoT security.

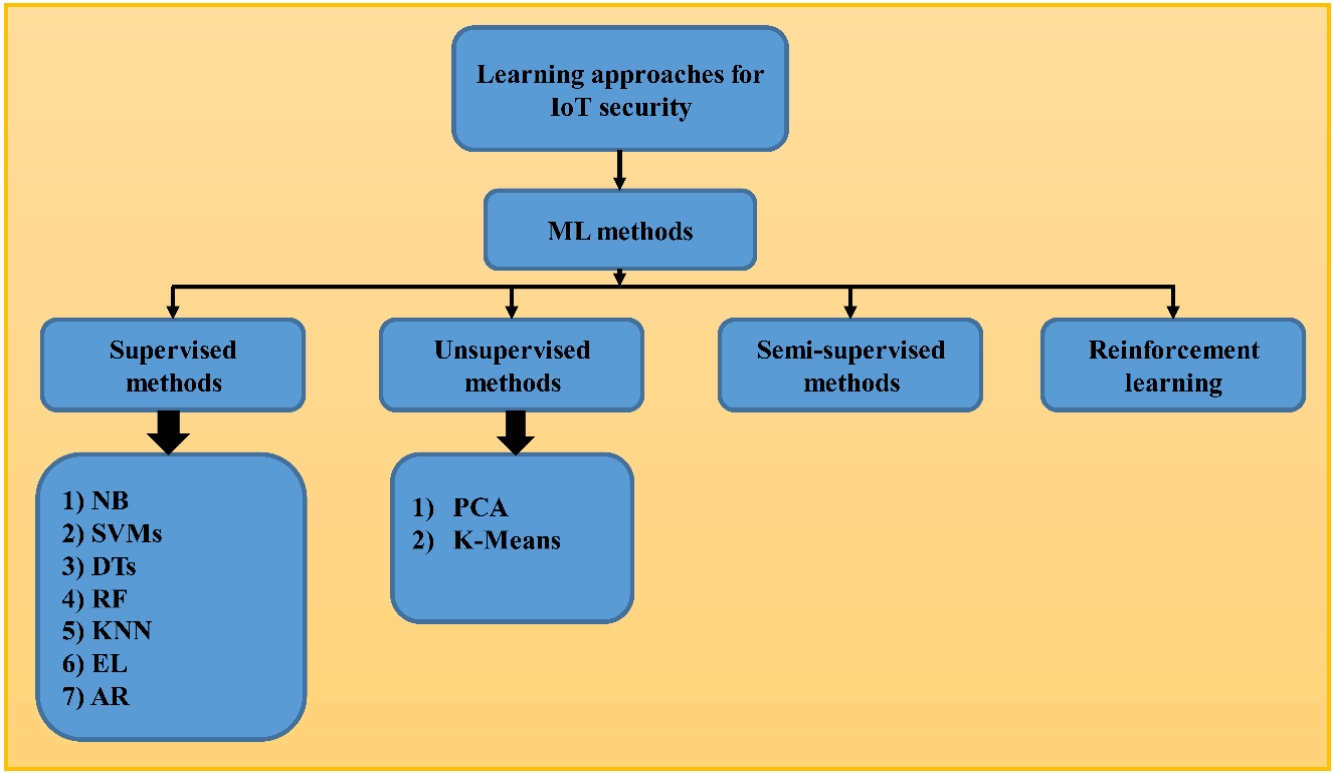

**Figure 2:** Various ML methods for IoT

### 6.1.1. Supervised Machine learning

We consider some traditional supervised ML techniques and merits, demerits, and applications in IoT for security enhancement.

#### 6.1.1.1. Bayesian theorem-based systems

Bayes' theorem illustrates the possibility of an event based on previous data associated with the event [90]. This is exemplified by the fact that DoS attacks are linked to network traffic information. Accordingly, Bayes' theorem is likely to evaluate the attack on network traffic by applying previous traffic facts aside from the above. A typical ML process, Naïve Bayes is a Bayes' theorem. As a result of its ease of use, it's sometimes called a supervised classifier.

NB estimates subsequent possibilities and applies Bayes' theorem. For specific feature sets, this theorem can forecast the likelihood. Such as, Naïve Bayes can be applied to categorize the traffic, including abnormal or normal. These features may be employed for the classification of traffic, for example, connection position flag, connection protocol (e.g., user datagram protocol (UDP) and transmission control protocol (TCP)), and connection duration are implemented or computed independently by this classifier. Despite that, features depend on each other. Each feature impacts to predict the probability in this classification method where the traffic is abnormal or normal. So, the Naïve was modified by hyperparameter tuning. This optimized model was applied for the detection of anomalies [91, 92] and network intrusion [93, 94]. The main benefits of these classifiers are ease, simplicity of implementation, the requirement of a small training sample, applicability to multi-class and binary classification [95], and toughness to inappropriate features [96].

The paper of [97] offered a method of intrusion detection to develop Bayes and PCA. It depicted that the Bayes classifier was more efficient than other classifier algorithms for detecting intrusion due to its rapid speed in the classification process. The intrusion of PCA can highly decrease the detection period. Next, the weight coefficient has been described to develop the PCA so that it can reduce the complexity of input data. The comparison between the detection time and rate using the traditional Bayesian method depicts that the technique introduced in this task is the best in intrusion detection. This work provided good accuracy. Besides, it also timely solved the requirement for detecting the intrusion of the network [97].

#### 6.1.2. Support Vector Machines (SVMs)

SVMs are mainly applied to evaluate data so that it is employed for classification and regression analysis. In the attributes of data, SVMs generate a separating hyperplane between the classes (two or more). The main target of the hyperplane minimizes the maximum adjacent sample features and distance between the hyperplane of each class. Each class has a maximum margin with minimum error [31, 98, 99]. Nonlinearity will occur if the research inputs cause the hyperplane to become confusable, necessitating the use of a kernel function to reform it. It is also challenging to use the appropriate kernel function in SVMs. With its high degree of precision, SVM is ideal for implementing network protection for IoT devices such as smart grid [100], ransomware [101], and intrusion [102].

SVMs have been utilized for classifying data by constructing a scattering hyperplane between two or more groups of data attributes that maximize the gap between the hyperplane and the class nearest sampling points [103, 104]. SVMs are well known for their broad spectrum of practical properties, but they are remarkably well suited to data sets with a limited number of sampling points [31, 98]. Theoretically, statistical learning [99] is designed for SVMs. Initially, they were designed to partition into a plane of two-dimensional composed of points of linearly independent data in various groups (i.e., abnormal or normal). This model will benefit from a good hyperplane to maximize distance by calculating the discrepancy between the closest points and the hyperplane in every class. It has benefitted from its ability and scalability to track intrusions in real-time and automatically change training tendencies. It has been commonly applied in a diversity of security applications, including intrusion detection [105-108], and is memory effective since they break data points using hyperplanes with $O(N^2)$ time complexity, where N is the sampling number [31, 98]. Research in [101] has created an Android malware identification tool to help protect IoT networks, as well as a linear SVM for its device in the context of the IoT. SVM's identification efficiency is superior to those of other computer algorithms such as Bayes naive, RF, and DT. SVM, on the other hand, outperformed the other ML algorithms. These findings support the robustness of SVM-based malware identification. However, more research is required to examine the efficiency of SVMs with enriched and attack scenarios of data sets generated in a variety of environments. In this case, it might be helpful to compare the efficiency of the SVM with that of deep learning algorithms like CNN. An SVM was previously used to protect an intelligent device, and an observational smart grid attack detector was tested [109]. This study found that ML algorithms including SVM, KNN, sparse logistic regression, and ensemble learning successfully identify unknown and known threats, outperforming traditional approaches used in intelligent grid applications. In another line of investigation, SVM was recently used to crack data encryption. The findings in [110, 111] demonstrate that ML techniques can be applied to hack cryptographic devices, and SVM outperforms conventional methods (such as template attacks).

**6.1.1.3. Decision Trees (DTs)**

The majority of DT-related classification methods are carried out by labeling samples based on their values of the attribute. Every vertex in a decision tree denotes one point, and every edge in the classification analysis represents a possible attribute for the vertex. Samples are categorized based on their attribute values and are classified from the root vertex [112, 113]. The feature that separates the training data optimally is referred to as the tree's initial vertex [114, 115]. Many approaches, such as the Gini index [116] and information benefit [117], are applied to deviate from training samples in search of the optimal function. The majority of DT approaches are split into two stages: classification (inference) and construction (induction) [118, 119]. DT is typically built by starting with an unoccupied tree and adding nodes and branches during the building (induction) phase. The feature that essentially divides the training samples is then called the tree's origin vertex. This role is chosen for a variety of reasons, including the value of experience. The idea is to delegate root nodes to reduce the intersection of groups in a training range, thus enhancing the discrimination efficiency of the classifier. Each sub-DT goes through the same process before all the leaves and associated groups are obtained. Following the development, new species are categorized with a collection of characteristics and an undefined class, beginning with the root nodes of the tree and progressing in the direction of the position values on the tree's inner nodes. This process is repeated before a leaf is collected. Lastly, the latest associated samples (such as expected classes) are determined [118]. Researchers summarized the crucial points for simplifying DT development in [118]. To begin, post- or pre-taking is used to reduce the tree's height. The state search space is then re-dimensioned. Third, the search algorithm has been updated. Following that, data attributes are minimized by disregarding or deleting unwanted features via the search procedure. To conclude, the tree's architecture is transformed into a data form, such as a law list. The table below summarizes the major drawbacks of DT-based approaches [118]. First, because of the house's design, they need a lot of room. Second, learning DT-based strategies is only simple when only a few DTs are involved. Certain structures, on the other hand, have a large number of trees and judgment nodes. The computational complexity of these applications, as well as the model underlying sample classification, is large. In defense applications such as intrusion prevention, DTs are used as the primary classifier or in combination with other master classifiers [120, 121]. In a previous study, for example, IoT devices were protected using a fog-based call system [122]. The thesis employed DT to examine network traffic in order to identify suspect traffic origins and, as a result, DDoS activity.

**6.1.1.4. Random forest (RF)**

The abbreviation "RF" refers to a supervised learning algorithm. Several DTs are built and combined in an RF to provide an accurate and reliable prediction model, resulting in improved overall results [123-127]. An RF is then made up of several trees that have been built at random and conditioned to vote for a certain class. The classification's final performance is determined by the most well-known class [123]. Since RF classification is mainly composed of DTs, these classification algorithms are very different. To begin, as the training set is fed into the network, the DTs generate a set of recommendations for classifying new data. RF constructs subsets of class voting rules that employ DTs; as a result, the designation contribution is the average vote, and RF is resistant to overfitting. Furthermore, RF eliminates the need to choose functions and permits a small range of input parameters [31]. However, in many real-time applications where the training data set is high, RF may be inefficient since it requires the creation of several DTs. Radio frequency techniques have been used to track network anomalies and intrusion detection [125, 128, 129]. In a work of [130], when narrow feature sets are applied to develop the device's application and reduce computational overhead to real-time classifications, ANN, KNN, SVM, and RF were learned to identify DDoS attacks on IoT systems, with RF being slightly stronger than other classificatory. RF was trained to recognize IoT interface groups from a white list of network traffic capabilities. The authors manually label and retrieve data from the network's seventeen IoT modules. These systems were divided into nine IoT device groups and applied to train a multi-class ranking using RF algorithms. According to the findings, ML algorithms are generally useful for correctly recognizing unauthorized IoT devices, especially RF [13, 55, 131]. Fig.3 depicts the basic architecture of RF.

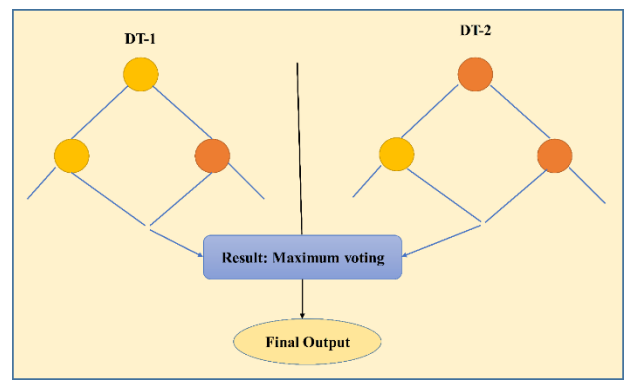

**Figure 3:** A basic architecture of RF

**6.1.1.5. k-Nearest neighbour (KNN)**

KNN is an ad hoc non-parametric approach. In KNN classifiers, Euclidean distance is often used as a distance metric [132-134]. Figure 4 shows how the KNN classification is used to characterize new input materials. The red circles in the diagram represent destructive behavior, while the green circles represent machine actions. The most recent sample (blue circle) must be labeled as benign or malicious. The classifier of KNN categorizes the novel example by voting a fixed number of times, i.e., the class of unknown samples is determined by KNN using a plurality vote of the closest neighbors. For example, if the classification of KNN is created on the nearest neighbor (where $k = 1$), Fig. 4 would classify the hidden sample as having normal behavior. (As the closest cycle is orange.) Since the two nearest circles are orange, the unknown specimen would be identified as having natural activity if the classification of KNN is trained on the two closest neighbors (where $k=2$ for normal behavior). If the classification of KNN has been trained on the four and three neighboring countries ($k = 4, k > 3$), the unknown sample class would be labeled as aggressive since the three and four circles closest to the unknown sample class are red circles (malicious behavior). Cross-validation is essential for evaluating the optimum rate of $k$ for a particular dataset. The KNN algorithm is an easy, high-performing classification algorithm on broad training data sets [135, 136], but the optimal $k$ value is still determined by the context. Choosing the optimum value of $k$ can also be a time-consuming and difficult operation. KNN classifiers have been used in network attack identification and detection of abnormalities [137-145]. In the area of the IoT, researchers of [146] recommended a paradigm for detecting R2L and U2R attacks. The algorithm decreased the dimensionality of the feature, allowing for two degrees of feature reduction to improve precision before introducing a model for a 2-tier classification based on KNN and NB classifiers. The suggested model performed admirably in detecting both attacks. Another study [147] developed a KNN focused on an intrusion detection method. The invention was intended to be used for node classification in a wireless sensor network (WSN). The proposed program was accurate and precise in detecting intrusions.

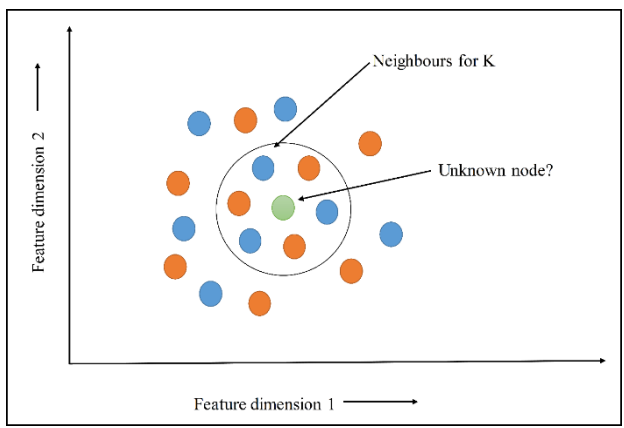

**Figure 4:** Principle of KNN

### 6.1.1.6. Ensemble learning (EL)

Ensemble learning (EL) is one of the most suitable methods of ML. EL integrates the outputs of various fundamental classification methods to achieve a single performance, increasing classification accuracy. This method attempts to integrate several multi-classifiers (such as heterogeneously or homogeneously) in order to arrive at a final output [148]. At the early stages of machine learning growth, each technique has advantages and accomplishments in particular implementations or datasets. Experiment comparisons in [149] revealed that the optimal style of learning varies depending on the application. The simple learning principle used to construct a classifier is determined by the data. Since the quality of data varies based on implementation, the best learning technique cannot be suitable for all applications. As a result, several classifiers have begun to be integrated to improve precision. EL employs a variety of learning techniques to eliminate inconsistency and is immune to overfitting. Combining several classifiers yields findings that go beyond the new range of theories; therefore, EL can respond well to a problem [150]. Due to the fact that EL is composed of several classifiers, an EL-based architecture has a higher time complexity than an EL-based scheme [151-154]. EL successfully detected intrusions, anomalies, and malware [155-158].

### 6.1.1.7. Association Rule (AR) algorithms

By analyzing the relationships between variables in a training data collection, AR algorithms [159] were used to characterize an unknown variable. Consider the variables X, Y, and Z in the P dataset. In order to analyze their links and to construct a model, an AR algorithm is intended to investigate the connections between these variables. The model then calculates the current sample class. AR algorithms describe regular sets of variables [31], which often coexist with vector collections in an attack. For instance, connections between IP/TCP variables and the attachment style using an AR were studied in a previous study [160]. The frequencies of various variables such as goal port, source IP, source port, and service name were analyzed in order to evaluate the attack form. The AR intrusion detection algorithm was successful, according to [161]. The researchers created a fuzzy rule-based model for intrusion detection, which resulted in a low false-positive rate and a high detection rate [161]. Nevertheless, compared to previous learning techniques, the increased reality is not widely utilized in IoT settings; more research is still recommended to determine if an approach to augmented reality may be combined or optimized with one more method to offer an appropriate solution for IoT protection. In practice, the following are the key disadvantages of the algorithms of AR. AR algorithms are difficult to deal with on a machine. If the frequency of the factors is reduced to an unmanageable level, association laws accumulate quickly. Although numerous productivity techniques have been developed, they are not often successful [162]. AR algorithms mostly focus on basic assumptions about variable relationships (direct relationships and occurrence). These assumptions are not always right, particularly when it comes to defending apps where attackers may try to mimic regular user behavior.

A study of [163] demonstrated how to reduce the time complexity to make it suitable for systems with minimal resources in hardware, for example, IoT devices. This work suggested an ensemble-based method that is lightweight and application-independent for identifying abnormalities in the Internet of Things. The recommended framework addresses two concerns: (1) automating as well as disseminating approaches in online learning for detecting device irregularities that are limited by resources, and (2) testing the proposed framework with actual evidence. According to the research, the ensemble approach produces each classification [163].

### 6.1.2. Unsupervised ML

In this subsection, we discuss the most frequently used unsupervised ML techniques (i.e., principal component analysis (PCA) and k-means clustering) and their applications, drawbacks, and advantages in IoT security.

Unsupervised learning occurs where the environment produces only inputs without regard for expected outcomes. It does not include branded data and can examine similarities between unlabeled data and classify them into distinct classes.

#### 6.1.2.1. Principal component analysis (PCA)

The PCA is a strategic reduction function that can be used to restrict a wide range of variables to a smaller list that maintains the bulk of the data. This approach reduces a large range of potentially associated features to some more minor uncorrelated features known as principal components [164-166]. There is an increasing order of variation among these components; the first is related with the most variance in the data, and the rest follow in order. Discard the components with the smallest variance. PCA is a fantastic solution in real-time scenarios. In contrast, there are several characteristics as well as it is difficult to see the connection and identify the correlation between each and every data. A large variety of accessible features also makes it quite difficult to narrow down the most important ones. Thus, the primary concept of PCA can be used to pick features for simultaneous detection of intrusion in IoT schemes. PCA helps speed up machine learning algorithms by removing linked variables that do not add to the algorithm's decision-making process. Moreover, to avoid the problem of overfitting, PCA reduces the number of dimensions in a dataset to a more manageable two [167,

168]. In this example, $n^2P + n^3 = O(n^2P + n^3)$ if each data point has P features and number of data is $n$.

The authors of the work [169] suggested a model that utilizes the reduction of features in PCA and classifiers such as KNN and softmax regression. According to the work of [169], these classifiers combining PCA resulted in a time as well as a computationally effective method. It can be used in IoT environments in real-time applications.

#### 6.1.2.2. K-Means clustering

Unsupervised ML is exemplified by K-Means clustering. This method seeks to find the data clusters, and $k$ represents the several clusters that the algorithm will generate. The process is used to allocate each data point iteratively to a cluster of $k$ according to the characteristics specified. Samples with identical characteristics can be included in each cluster. The $k$-means algorithm results in iterative refining. Two inputs are needed for the algorithm: several clusters ($k$) and a dataset with the specification of each example in the dataset. According to the estimate of the $k$ centroids, each sample is allocated to the cluster centroid nearest to it, depending on the distance from the Euclid. Secondly, after all, data samples have been allocated to a given cluster, the cluster centroids would be reevaluated, applying the mean of all samples allocated to that cluster. The algorithm repeats these calculations until no samples can be used to modify the clusters [170, 171]. The preceding are the key disadvantages of clustering in $k$-means. To begin, the user must enter $k$. Second, this algorithm is based on the premise that each spherical cluster has nearly equivalent sample counts. The algorithms of $k$-means may be used to detect irregularities by contrasting normal and odd behavior features [172, 173]. The authors of [174] suggested an anomaly recognition scheme focused on DT and $k$-means (such as DT, C4.5 algorithm). Nevertheless, k-means performed less well than directed learning approaches, especially when detecting established attacks [175]. When it is difficult to obtain labeled outcomes, unsupervised algorithms are usually a safe bet. Nonetheless, clustering methods in general, especially $k$-means, are still in their infancy for stable IoT structures and should be studied further. ML techniques that are not regulated have many applications in IoT network security. K-means clustering, for example, has been used to protect WSNs by detecting intrusions [176]. In research on Sybil recognition in industrial WSNs, a kernel-based scheme was suggested for clustering channel vectors to distinguish Sybil from standard sensors [177]. A clustering algorithm demonstrated the possibility of anonymizing private data in an IoT scheme [178]. In implementing anonymized data algorithms, the usage of clustering would significantly improve the protection of data sharing [178].

### 6.1.3. Semi-supervised ML

Regulated machine learning is the most widely used ML process, and it derives its information from the training period on labeled data. To begin, developing predictive models from labeled data takes time, money, human interaction, and expertise. On the other hand, unsupervised learning that operates on unlabeled data frequently has an exploratory component (i.e., compression, clustering). Thus, the researchers hope that by implementing a semi-supervised approach, they would be able to solve the problem of producing vast quantities of labeled data required for training algorithms in supervised machine learning by augmenting unlabeled data [179, 180]. Accordingly, semi-supervised learning deploys both classified and unlabeled input to train a machine learning classifier. Nevertheless, while semi-supervised learning can seem to be a suitable answer to the problems associated with both managed and unsupervised approaches, it can fall short of the prediction precision attained by the algorithm in supervised machine learning. As a consequence, limited studies have looked into the use of semi-supervised methods for protection in IoT. Such as, the authors of [181] described a semi- supervised multi-layer clustering (SMLC) technique for detecting and preventing intrusion in the network. SMLC has demonstrated the ability to learn from incompletely labeled instances while also gaining recognition performance comparable to managed machine learning for detection and avoidance systems [182].

### 6.1.4. Reinforcement learning (RL) approaches

One of the first topics that come to mind is learning from one's own surroundings. People naturally begin their education by engaging with their surroundings. RL is driven by neuroscientific and psychological observations of animal behavior and mechanisms that enable agents to have a more significant effect on their environment [183-185]. RL shows people how to better map conditions to actions in order to maximize rewards [184]. The agent does not realize which actions to do ahead of time and must determine the acts that have the most significant benefit by assessment or error. The features' trial and 'error' are the key characteristics of RL. As a result, the agent continues to gain expertise in order to maximize the benefits. RL has been used to address a variety of IoT-related problems. The work of [186, 187] suggested a broadband, autonomous cognitive radio anti-jamming system with an emphasis on learning enhancement (WACRs). Data was used in [186] to differentiate between the swinging jammer signal, unintended interference, and former WACRs; reinforcement learning was then applied to reliably learn a selection technique of subband to stop the jammer signal as well as interfere with previous WACRs. Likewise, the authors of [187] discovered how to efficiently prevent jamming attacks from hundreds of MHz of spectrum in real life using an enhanced learning system based on Q-learning. Another work [188] used similar strengthening training to develop a cognitive radiation anti-jamming method, which was paired with deep CNN to increase the performance of RL across a wide range of frequency sources. A related method was suggested in [188] to tackle aggressive jamming using a rigorous learning approach; the findings showed that RL was an appropriate tool for modeling aggressive jamming schemes.

### 6.1.5. Applications of ML in IoT Security

In data analysis, semi-supervised and supervised methods are used, whereas comparative and decision-making are favored for reinforcement. The essence of accessible data dictates the categorization and methods used by ML. Supervised learning is used to determine the form of input data and the desired outcomes (labels). In this case, the machine was taught to only map the inputs to the necessary outputs. Regression and classification are supervised learning processes that utilize constant data regression and discrete data classification, respectively. Many regression methods, such as polynomial regression, linear regression, and SVR are widely used [55, 189, 190]. In contrast, classification employs distinct production qualities (class labels). K-nearest neighbor, SVM, and logistical regression are provided by commonly used classification algorithms. Certain architectures, such as neural networks, may be applied for both regression and classification. Where the results are not well-defined, and the process must search within the raw data framework, unchecked learning approaches are applied to teach the algorithm. Clustering is a form of unattended learning in which items are clustered based on similar parameters, i.e., K-means clustering. The accuracy of predictive analytics is determined by how effectively master learning utilizes historical data to create models and how well future values are estimated. Algorithms such as Naive Bayes, and SVM are applied in predictive modeling. The one disadvantage of simple machine learning approaches is that they need a large amount of data for model testing. The learned model is then used to approximate or interpret real-world application performance. However, it should be remembered that the whole procedure would not capture the whole spectrum of data and resources. To address the shortcomings of machine learning methods, DL techniques have been deployed. DL can handle large amount of data, and its algorithms are adaptive as data volume increases, benefiting model training and possibly improving prediction precision. DL extracts high-level functionality and associated connections from input data in a complex and hierarchical manner. The majority of IoT implementations produce outcomes without labeling or with semi-labeling. DL may use unlabeled data to identify valuable trends in an unattended manner. Conventional machine learning algorithms are only effective when there is a large number of labeled data available [46].

The work of [182] suggested a semi-supervised learning IoT attack prediction mechanism. The proposed system is based on the algorithm of the Extremely Learning Machine (ELM), using Fuzzy C-Means (FCM) [191] approaches collectively known as Extreme Learning Machine(ELM)-based Semi-Supervised Fuzzy C-Means (ESFCM). Similarly, ESFCM is used in fog infrastructure. One characteristic of ESFCM is that it operates with marked directories, improving the rate of detection of threats transmitted. Although the detection performance of ESFCM is lower than those of the two former DL mechanisms, it outperforms traditional attack detection machine learning algorithms. Nevertheless, the semi-supervised learning process incorporates the benefits and effectiveness of managed and uncontrolled learning. The IoT has a multitude of flavors, from body area networks to sophisticated key business facilities such as a smart grid. At the same time, identifying attacks on these infrastructures is important. In an intelligent grid, for example, steps are critical and must be recovered authentically and without alteration as a result of an attack. The authors of [109] conducted a comprehensive study in this direction to explore various algorithms in ML for attack recognition in smart grids. The researchers studied the function of space fusion, semi-monitored learning, online learning, and supervised learning algorithms. The authors considered the effectiveness of online approaches for detecting attacks in real-time by concentrating on their numerical complexity, that is usually below that of batch learning algorithms [109]. In contrast, all families of the above algorithms performed fairly well.

## 6.2. DL in IoT Security

Deep learning has developed a key topic for study in recent years in IoT systems [14, 192, 193]. The main advantage of deep learning over classical machine learning is its higher effectiveness for big datasets. Many IoT schemes produce huge quantities of data; therefore, for those systems, DL methods are well-suited. Furthermore, DL dynamically generates dynamic data representations [194]. The IoT ecosystem can be connected in-depth with DL methods [195]. Deep connection is a unified protocol that facilitates automated communication between computers and applications linked to the Internet of Things. For example, in an intelligent home, IoT devices automatically talk to each other to form a completely intelligent home [14]. DL approaches use a computational paradigm that integrates many layers to learn different degrees of abstraction in data structures. Compared to traditional ML approaches, DL techniques have greatly enhanced state-of-art methods [196, 197]. DL is a subfield of ML using various non-linear layers of computation to abstract and turns discriminatory or generative pattern analysis functions. Since DL methods may catch hierarchical images in deep architecture, they often refer to themselves as hierarchical methods of learning. The operational theory of DL is motivated by the interpretation of impulses by the human neurons and brain. Deep networks are used to include unsupervised, supervised learning and a combination of these two learning forms, such as deep hybrid learning. This section discusses the most commonly used deep learning algorithms. Fig. 5 depicts several DL classifiers for IoT security.

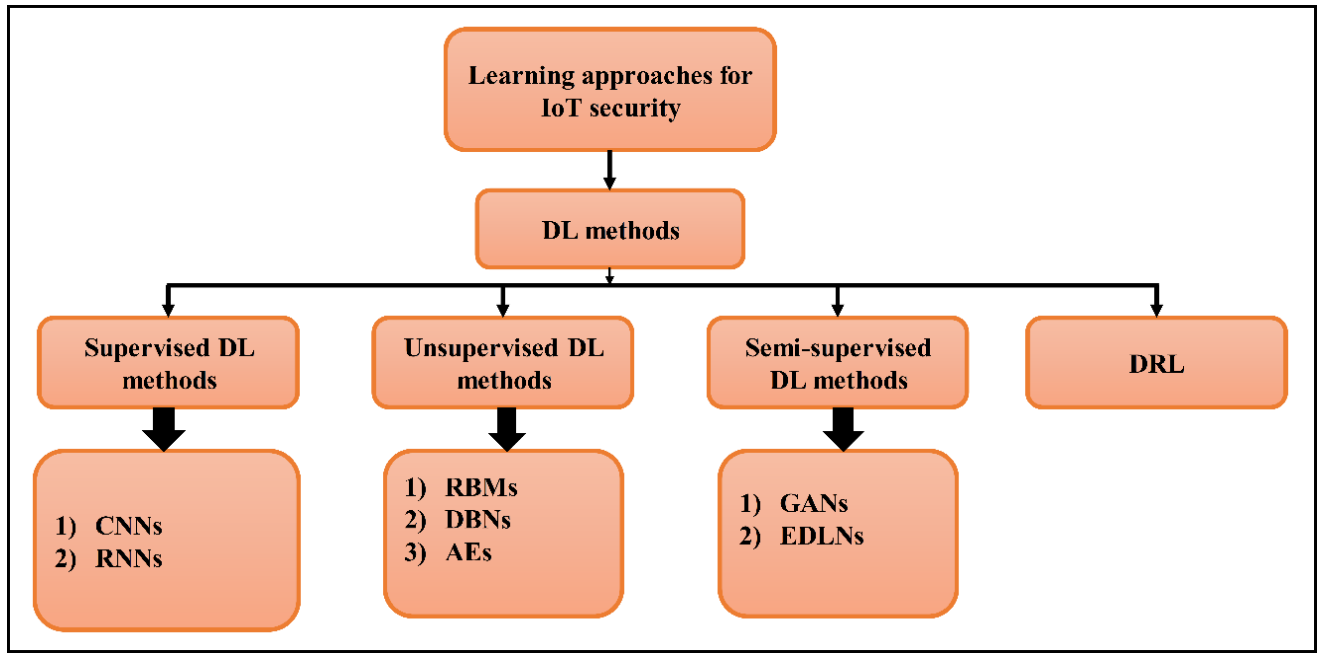

**Figure 5:** Various DL methods for IoT

**6.2.1. Supervised DL**

This subsection discusses the most often used controlled DL methods. Recurrent neural networks (RNNs) and convolutional neural networks (CNNs) are two types of discriminative DL algorithms.

**6.2.1.1. Convolutional neural networks (CNNs)**

CNNs are created in order to decrease the number of data parameters used in typical neural artificial networks (ANN). To minimize data parameters, three terms are used: sparse relationships, parameter sharing, and fair distribution [198, 199]. Reduced layer-to-layer relationships improve CNN scalability and complexity. A CNN comprises two types of layers: convolution and convergence. Convolutional layers combine data parameters by using a variety of analogous filters (kernels) [200]. The pooling layers' sample reduces the size of the following layers through average pooling or peak pooling. The top pooling algorithm splits the input by non-overlapping clusters, selects the highest value in the previous layer for each cluster [201, 202], and then combines the values of each cluster in the previous layer with the average pooling algorithm. The activation device is another critical layer of the CNN; each vector in the feature space has a function in non-linear activation. The ReLU is chosen since it contains nodes with the activation property $f(x) = \max(0, x)$ [203]. Fig. 6 depicts how CNN works when IoT protection is extended. The biggest drawback of CNN is that it is widely used in deep learning strategies. It also enables high-performance automatic learning of raw data functions. However, because CNNs have a high machine cost, resource-constrained devices that support onboard security schemes are difficult to deploy. Distributed architectures should solve this problem. In this design, a light deep neural network (DNN) is introduced and equipped, but the algorithm is entirely trained in the strongly categorized neuron [204], with a subgroup of similar output groups on board. The advancement of CNNs is primarily targeted at image detection. As a result of their wide use, CNNs are used to create accurate and consistent models of Image ID and Classification for massive public image databases such as ImageNet [205, 206]. CNNs also display their worth in a variety of other applications. According to one research [207], a CNN-based IoT protection malware identification framework for Android could be created. CNN is used to acquire major malware identification characteristics from raw data.

The main argument for CNN usage is that sufficient functionality is taught concurrently with classification, eliminating the extraction step needed for conventional machine learning and producing a full model [207]. However, attackers may use the robust learning capacity of CNNs as a weapon. A past study [208] depicted that a CNN algorithm is efficiently capable of breaking cryptographic applications.

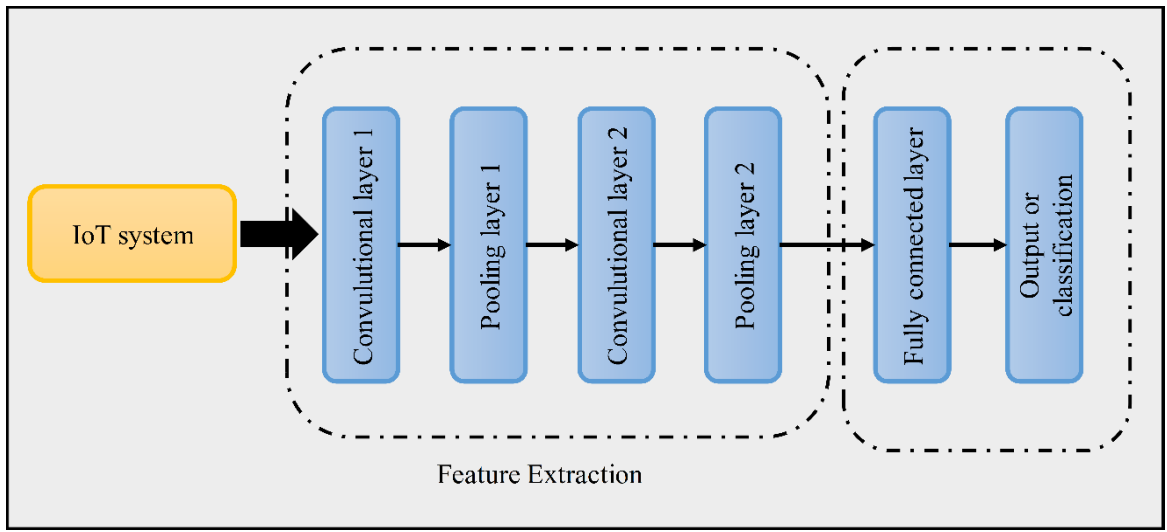

**Figure 6:** CNN for IoT security

**6.2.1.2. Recurrent neural networks (RNNs)**

Recurrent neural networks (RNNs) are a crucial type of deep learning algorithm. RNNs are suggested to deal with temporary results. The provision of current performance is dependent on an interpretation of the similarities between several previous examples in many implementations. The neural network performance is therefore determined by its current and previous inputs. Since the input and output layers continue to be distinct, a feed-forward NN is unavoidable in this design [209]. When the backpropagation algorithm was created, it was primarily used for training RNNs [196, 210, 211]. RNNs are recommended for appliances that need sequential inputs (e.g., sensor, text, and voice data) [210]. An RNN has a sequential data storage layer and knows many elements of the recurrent cells secret units [212]. In addition to the network details, the secret units are updated and continuously changed to represent the network's current state. The RNN forecasts that the previously hidden state would be activated when working on the currently hidden state. RNNs are applied because they can efficiently manage sequential data. This skill is useful for a variety of activities, including the identification of dangers where the patterns of the danger are time-based. As a result, recurring associations may be used to strengthen neural networks and uncover prominent behavioral trends. The main disadvantage to RNNs is that gradients disappear or burst [213]. RNNs and their derivatives have outperformed in a variety of sequential data implementations, including speech recognition and translation [214-216]. Furthermore, RNNs may be used to secure IoT computers. IoT networks gather massive volume of sequential data from a variety of sources, including network traffic patterns, and are required to identify a variety of possible network attacks. An earlier study [217] checked the feasibility of RNN analysis of network traffic activity for identifying possible attacks (malignant action) and validated the RNN utility for network traffic classification to effectively detect malignant behavior. RNNs are also a viable solution in real-world situations. The investigation of RNNs and their variants is critical for improving IoT system protection, especially against serial-based attacks.

### 6.2.2. Unsupervised DL

We discuss the most often used unsupervised DL techniques, including deep restricted Boltzmann machines (RBMs), deep belief networks (DBN), and autoencoders (AEs) in this subsection.

#### 6.2.2.1. Restricted Boltzmann machines (RBMs)

Uncontrolled RBMs are deep generative models [218]. An RBM is an entirely undirected model in which nodes are not bound within the same sheet. RBMs are divided into two types of layers: exposed and unseen. The input is used in the visible layer, while the opaque layer comprises latent variables of various levels. RBMs accumulate data features in a hierarchical manner and are applied as hidden variables in the subsequent layer to record features in the initial layer. The study in [219] presented a model for detecting network anomalies that overcome difficulties in developing this model. This issue includes creating labeled data needed for effective model testing due to the multipart erratic nature of network traffic data collection. The second issue is that irregular behavior often evolves over time. The model can then be continually modified, allowing new attack types to be recognized and anomalies in a variety of network environments to be observed. The researchers recommended a learning model focused on a discriminatory RBM in [219], which they chose because of its ability to combine generative models with enough classification precision to identify a half-controlled network anomaly even though training figures are incomplete. However, their experimental findings revealed that the discriminative RBM classification exactness was reduced when measured on a different network dataset than the data used to train the classifier. More research is needed to determine if an exception to a classifier in a variety of network contexts may be extended. A single RBM may only display a limited range of functions. However, RBM may be used to build DBN in significant ways by piling two or more RBMs. The part that follows goes into more detail on this strategy.

#### 6.2.2.2. Deep belief networks (DBNs)

The generative approaches for DBNs are taken into account [220]. A DBN comprises stacked, layer-free RBMs that run greedy workouts in a stable, unmonitored setting. In a DBN, teaching is done layer by layer, with each layer being an RBM trained on the previously trained layer [212]. The initial characteristics [212] are learned using a greedy layer-specific unmonitored technique during the pre-training level. The top layer is finished using a Softmax layer during the fineness phase [217]. DBNs have been successfully used to detect malware attacks. An earlier study [221] suggested a method for the security of mobile edge computers by the use of a profound learning technique to detect malicious attacks. DBN was used for automatic identification. Compared to machine learning-based algorithms, the proposed DBN model significantly improved malware detection precision [221]. This finding demonstrated that deep learning approaches, specifically DBNs, outperformed conventional manual malware identification feature engineering methods. An EA was combined with a malware detection method utilizing DBN in a recent study [222]. By non-linear projection, an algorithm in AE DL was applied to reduce the dimensionality of data and delete only the significant functions. DBNs are unregulated learning methods that are trained on unlabeled data to reflect significant features. Although DBNs use conflicting convergence to minimize processing time, they do not function for onboard computers that have limited resources.

#### 6.2.2.3. Deep autoencoders (AEs)

A deep AE is an unsupervised learning neural network that has been learned to replicate its input to output. A secret layer h specifies a code used to describe the input in an AE [198]. An AE neural network is split into two parts: the encoder function $h = f(x)$ and the decoder function $r = g(h)$, which tries to replicate the data. The encoder receives the feedback and transforms it into an abstraction known as a code. Following that, the decoder obtains the built text, which was originally generated to reflect the data, in order to reconstruct the original input. The learning phase in AEs can be completed with the least amount of reconstruction error [45, 223]. However, AEs cannot be taught to precisely reproduce the feedback. AEs are also constrained by being able to provide an approximate reproduction only by merely copying inputs similar to the training outcomes. The model must prioritize which input characteristics should be copied; therefore, useful data features are continually learned [198]. AEs have the ability to be useful for function extraction. In contrast, AEs provide a significant amount of computing time. Although AEs can learn to collect the characteristics of the training data effectively, if the training dataset does not match the test data, they can confuse the learning process instead of reflecting the data collection. In [224], Network-based AEs have been used to recognize ransomware, and AEs have learned the latent representation of a dynamic function set, focusing on the cyber system's vector. The AEs outperformed the standard ML algorithms such as KNN and SVM in terms of detection efficiency [224]. In other research [222], A DBN was added to create a malware detection method that was then used to reduce data dimensions by non-linear mapping so only the important features could be eliminated. Next, the algorithm in DBN learning was learned to identify malicious code.

**6.2.3. Semi-supervised or hybrid DL**

This section discusses the most traditional deep hybrid learning approaches. Among the hybrid DL approaches are generative adversarial networks (GANs) and network communities (EDLNs).

**6.2.3.1. Generative adversarial networks (GANs)**

GANs, which were recently pioneered by [225], is already an exciting platform for deeper learning. As seen in Fig. 7, two models, both generative and discriminatory, are trained concurrently by a GAN approach using an opposed mechanism. The generative model learns the data distribution and outputs data testing, while the discriminatory model predicts the probability of the results from the evaluation rather than the generative model. The objective of training the generative model is to increase the probability that it is wrongly classified by the discriminatory model [225, 226]. By changing the sample dataset, each phase trains the generative model to fool the discriminator. The model serves as a generator. In this regard, the discriminator is given many individual data samples from the training array and the generator samples. The discriminator is used to distinguish between actual and fake specimens (from the training data set). Using incorrectly labelled samples, the outputs of unequal and generative models were quantified. The following edition's versions are then revised. The performance discriminative model aids in the generation of samples for the next iteration while optimizing samples for the next iteration [45]. GANs were recently added to IoT protection. In [227], architecture was built to protect the cyber field of IoT networks, including the training of in-depth learning algorithms to distinguish between ordinary and abnormal computing. GAN algorithms were used in the suggested architecture for the preliminary analysis, and the test results demonstrated the architecture's efficacy in detecting suspicious system activity [227].

Since GANs can learn various attack scenarios and produce a zero-day attack-like sample, they can provide algorithms with samples that are not available from current attacks. GANs are well-suited for semi-supervised classification instruction. Since GANs are not sequentially needed to produce several accesses in the samples, they can produce samples faster than fully transparent DBNs. In GANs, sampling requires only one stage in the model, while RBMs need an unknown number of Markov chain iterations [225, 228]. GAN teaching, on the other hand, is risky and challenging. A GAN cannot be used to produce different data such as text [225, 228].

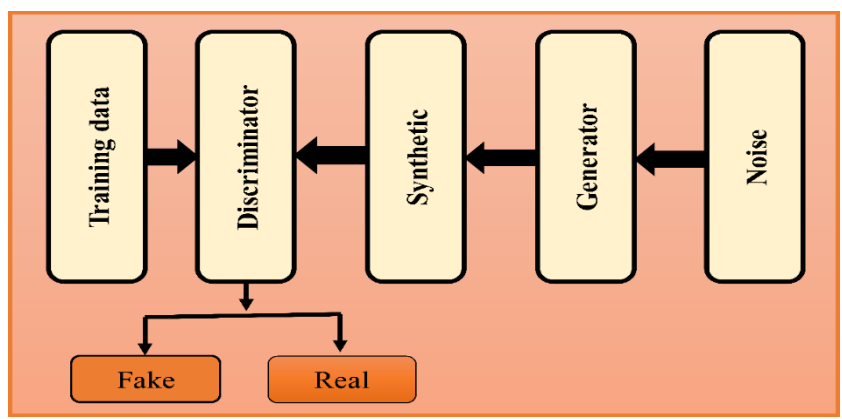

**Figure 7:** Working diagram of GAN

### 6.2.3.2. Ensemble of DL networks (EDLNs)

It is possible to create EDLNs using hybrid or discriminative and generative models. An emerging algorithm in deep learning, ensemble learning (EL) employs a variety of classification approaches to improve its performance [197, 229-231]. Homogeneous or heterogeneous multi-classifiers are often used to get an accurate result. Most issues can be solved using the various methods used by EL. When compared to other single classifier methods, EL takes a long time. Anomaly detection, virus detection, and intrusion detection are all typical applications for EL [156, 163, 232]. Combining DL classifiers may assist in achieving model variety, increasing model performance, and extending model generalization using the EDL approach. EDL's key drawback is that the system's temporal complexity may be significantly enhanced.

Due to numerous neural networks' training, assembling DNNs does not seem to be a feasible alternative due to the potential for a significant rise in computing cost. Deep networks may be trained on high-performance hardware using GPU acceleration over the course of many weeks. Explicit/Implicit ensembles achieve the conflicting objective of training a single model. It acts as an ensemble of training several neural networks apart from incurring extra or with as little additional cost as feasible. In this instance, the training time of an ensemble is identical to that of a single model. In implicit ensembles, model parameters are shared, and the model averaging of ensemble models is approximated by a single, unthinned network during test times. In explicit ensembles, however, model parameters are not shared. The ensemble output is determined by combining the predictions of the ensemble models using various methods such as majority voting, averaging, etc. Dropout [233] generates an ensemble network by arbitrarily removing hidden nodes from the network during training. During the testing period, all nodes are operational. Dropout offers network regularization to prevent overfitting and adds sparsity to the output vectors. Training an exponential number of models with standard weights and providing an implicit ensemble of networks during testing reduces overfitting. Randomly dropping the units prevents coadaptation by making the existence of a specific unit unpredictable. The network with dropout takes 2 to 3 times longer to train than a regular neural network. Therefore, a suitable balance must be struck between the training duration of the network and overfitting. DropOut is described in detail in DropConnect [234]. In contrast to DropOut, which eliminates each output unit, DropConnect randomly eliminates each connection, introducing sparsity in the model's weight parameters. Similar to DropOut, DropConnect generates an implicit ensemble at test time by deleting connections (setting their weights to zero) during training. Both DropConnect and DropOut have a lengthy training period. To address this issue, deep networks with stochastic depth [235] sought to minimize the network depth during training while maintaining it during testing. Stochastic depth is an enhancement on ResNet [236] in which residual blocks are eliminated at random during training, and transformation block connections are bypassed via skip connections. Swapout [237] extends DropOut and stochastic depth.

Many deep learning algorithms will be able to outperform separately deployed algorithms if they operate together. EDLNs can be generated by integrating generative, discriminatory, and hybrid frameworks. EDLNs are frequently applied to address complicated problems, including uncertainty and a wide variety of dimensions. An EDLN is a stacked collection of heterogeneous (classifiers from separate families) or homogeneous (classifications from the similar family) classifications. They're used to boost variability, precision, efficiency, and widespread [238]. For example, the authors of [239] use a sparse auto-encoder (SAE) to extract attributes and the softmax activation with a regression layer to create classifiers. The evolutionary findings indicate that we could obtain a higher level of efficacy than in previous studies utilizing a semi-supervised intrusion detection technique. EDLNs have proven to be surprisingly successful in a variety of applications, such as activity detection for humans, but their usage in IoT protection necessitates additional testing, especially the ability to deploy light homogeneous or heterogeneous graders in a dispersed setting in order to increase IoT security system accuracy and efficiency and address machine challenges.

### 6.2.4. Deep Reinforcement Learning (DRL)

Reinforcement Learning (RL) has been introduced as an efficient technique of improving a learning agent's methods and determining the optimal solution through evaluating and failing to achieve the best long-term goal without previous environmental awareness [240]. RL is a kind of ML. An agent learns how its actions affect the environment via trial and error in RL. After each action, it calculates the reward and then proceeds to the next state [241]. It is possible to utilize RL to tackle very complicated issues that are intractable with traditional methods since it focuses on long-term outcomes. While no training dataset is available, it is the best option for learning by experience and contact with the environment. Moreover, Real-world issues are assumed to be Markovian models in RL (which is not the case in reality), where the agent is in a state $s$ at all times, performs action $a$, and gets an integer reward before changing states $s'$ according to the dynamics of the environment, which is represented by $p(s'|s,a)$ in this case. If an agent is trying to maximize its returns, it will try to learn a policy based on its observations or mapping of those observations to actions (expected sum of rewards). Unlike in optimal control, in reinforcement learning, the algorithm only has access to the dynamics $p(s'|s, a)$ via sampling. DRL techniques use deep learning to solve Markov decision processes (MDPs) like these, frequently modeling the policy $\pi(a|s)$ or other learnt functions as a neural network and building specific algorithms that operate well in this environment.

Deep Reinforcement Learning (DRL) approaches, for example, a deep Q-network (DQN), have been used in a variety of mobile edge computing applications to solve high dimension problems while still providing scalability and download performance [242]. The Deep Q network [219] is a recently used strengthening tool. Many proposed deep Q network upgrades have been proposed, such as dual Q-learning [243], continuous monitoring by deep RL [244], and priority replay of knowledge [245]. The curse of dimensionality restricts its application to real-world systems, making it inapplicable. In addition, it requires a lot of data and computing to run. As RL has a number of drawbacks, it is often used in conjunction with other ML approaches. DRL is a standard combination of RL and DL.

The authors of [242] investigated access management and download for mobile edge cloud computing, as well as the integration of blockchain and DRL in application schemes in IoT networks. In another area of research, DRL was used to protect cyber security. In [246], the authors investigated many DRL cyber protection approaches, such as DRL-based cyber-physical network security mechanistic mechanisms, automatic intrusion prevention tactics, and multi-agent cyberattack mitigation model DRL-based game theory. It may be a step forward to investigate these approaches within the context of IoT.

**6.2.5. Application of DL in IoT Security**

DL methods have been established for signal authentication in IoT settings [247, 248]. In order to derive arrays of stochastic properties from IoT device signals, Ferdowsi et al. [247] proposed an LSTM architecture. The properties of cyber-attack control in IoT systems were then watermarked, and the complex extraction of functions often helped detect eavesdropping attacks. This solution would not, however, apply to very large IoT settings as it is prohibitively difficult to authenticate all IoT devices in a centralized cloud service. In order to solve this challenge, we have merged the LSTM-based signal authentication process with game theory methods for a mixed Nash balance (NE) strategy. Although this approach is capable of handling many IoT modules, it is also highly dynamic and not suitable for IoT environments. Authentication in DL-based systems using LSTM is also recommended for IoT environments [249] since it has been designed for device recognition to be resilient to signal imperfections. Nevertheless, the method is effective for system recognition and cannot notice other severe attacks. DNN has been updated by utilizing intrusion detection systems (IDSs) to classify substantial attacks in the IoT environment [250]. In order to validate DNN performance, cross-validations, cross-validations, and subsampling were used, and parameters in DNN were reset with a method for grid quest. While DNN performed well on many datasets, including those with imbalanced and distorted results, it was concluded that the determination of learning parameters for the grid search took more time. DL approaches have been used to defend Social IoT (SIoT) [251], with DL methods spreading via fog nodes to enable the identification of a distributed threat. The DL solution detects the following kinds of attacks with SIoT: sample, DoS, U2R and R2L. The analysis exhibited that the DL solution performs other machine learning strategies, while more research is required on network intrusion detection based on payload. A deep learning methodology has been developed for dense random networks [252] to identify network attacks. The dense network-oriented DL approach focused on these metrics was used to observe attacks on ioT gateways. The main drawback of this solution is that the parameter setting of the dense random network is needed since the non-optimal parameters are not properly classified. In IoT healthcare environments, protection and privacy have been maintained by the use of layering-based deep-Q networks [253] that access control, enable authentication and mitigate intermediate attacks in the IoT environment. Packet features including protocol, IP address, post number file type, frame number and frame length were initially gathered and stored in a local database. Deep-Q networks are used to identify medical data according to the derived functionality and the classification is achieved using the packet functionality since the extraction of optimal features from data packets achieves a higher degree of accuracy. A deep Eigen space learning approach has been suggested by the Battlefield Things (IoBT) Internet for malware detection, [72] in which the Operational Code (OpCode) device sequence is used to classify malware. The OpCodes interface was translated to vector sequences, and a deep Eigen spatial learning technology was used to differentiate between beneficial and malicious programs. The entire process started with the development and classification of a graph for each sample. This technique includes considerable computation, which limits its performance when handling big datasets. Identification of DDoS attacks was conducted in IoT systems using IoT-specific network features such as tiny endpoints and day-to-day packet cycles [130]. A network-dependent method was also suggested for detecting IoT botnets on the basis of deep autoencoders [254]. Five separate cycles of N-BaIoT functions, such as packet jitter, packet count, and packet size, have been obtained. This approach only incorporates static characteristics that restrict the self-encoder efficiency and require optimum feature selection to increase the accuracy of the autoencoder. Bidirectional LSTM-RNN (BLSTM-RNN) was applied in [255] to detect IoT botnets, and word embedding was used for text identification and the transformation of attack packets to integer sizes. Although BLSTM-RNN detects botnets with a limited number of attack vectors, it does not work when the count of attacking vectors increases. Autoencoder was introduced in a fog-enabled IoT framework for stacked unmonitored deep learning approaches [61], and architecture of fog-based IoT was built to improve attack detection latency and scalability [256]. While deep learning models exceed shallow attack detection learning algorithms, stacked auto encoders increase the computer's runtime and its accuracy. In Industrial IoT (IIoT), the systems of network intrusion detection are applied to secure a network against a range of security threats [256]. For intrusion detection, a deep autoencoder and a deep feed forward neural network are deployed, and feature transformation and normalization are used in the overall process. This method requires more research to establish how different intrusion mitigation protocols should be handled in the IIoT [257]. Significant routing attacks have been defined and evaluated with the following characteristics, such as decreased rating, hello flood, and change of the version number: average receipt speed, packet number, transmission rate, average transmission period, packet counting and overall transfer time. Without accessibility, detection effectiveness and the possibility of detecting such extreme IoT attacks were impaired. The works of [251] proposed a scheme to identify distributed attacks on IoTs and contrasted their effectiveness with conventional machine learning methods and distribution networks as well as the centralized approach of detection. The results were also tested for two and four-class grades; the two-class grades included regular and attacks, while the four-class grades included U2R, DoS, probe, and natural and remote-to-local grades. The utility of distributed detection has been evaluated on a range of network training engines, relative to standard IoT learning approaches for DL intrusion detection. The findings showed that the distributed architecture exceeded the central structure with an accuracy of detection varying from 96% to 99%. Furthermore, results exhibited that DL approaches were more precise than conventional learning methods and had a lower false alarm rate of 0.85 percent compared to machine learning methods. DL had a 99.27% recall rate and an average 96.5% recall rate, while ML had a 97.50% recording rate and an average recording rate of 93.66%. These results show that deep learning techniques in distributed IoT environments are highly likely to identify cyberattacks. Experiment results show that distributed sensing systems outperform hierarchical methods in detecting cyber threats because they exchange variables, preventing the formation of local minima during planning. The research can be covered by a comparison of distributed approaches to deep learning to different traditional methods of learning on different data sets. Techniques for the analysis of the network load data to identify intrusions through key trends may also be further explored.

## 7. Solutions to IoT Threat using DL or ML algorithms

Privacy and security concerns can be mitigated in many ways. In section 3, we described several types of threats in IoT. There were no solutions and discussions about how to ensure security or privacy in the IoT system. Hence, we concentrate on recent works suggesting privacy and security-preserving methods for the IoTs in this Section. We illustrate the solutions suggested by DL or ML algorithms as a tool for ensuring privacy and security.

A security program is a collection of policies and procedures designed to safeguard an organization's most sensitive data and assets. Instead than concentrating on people's private details, it highlights statistics and other facts. On the other hand, passwords, login information, and other sensitive data are the primary targets of privacy programs.

Safeguarding privacy, maintaining data and information's integrity, and making sure it's readily accessible are the three pillars of security. The right to the confidentiality of one's own and one's employer's private data is a cornerstone of privacy. Security measures may help provide some level of confidentiality, and the secrecy of credentials and access to data is essential to a robust security framework.

Machine learning (ML) is a data processing technology that is applied in all frameworks' data processing pipelines. For instance, a machine learning model may assess data flow into a network in order to get an up-to-date decision. Poisoning or exploratory attacks on the input data from the source to the IoT nodes, as well as on the IoT nodes to the ML model, are possible. Inversion and integrity attacks are feasible on the output [258]. As a result, the privacy and security of a system cannot be compromised simultaneously.

### 7.1. Security Solutions

On the basis of DL and ML methods, several suggested security measures are shown in Table 1. According to the work of [62], fog computing decreased the threat of spying in communications as well as attacks in MiTM by limiting interaction to IoT devices in close proximity during flooding attacks. Based on this, they implemented their model by applying the long short term memory (LSTM) method, which can keep track of historical data. In order to compare their findings with LR's, they used the ISCX2012 dataset. It included 71,617 instances of DoS attacks and 440,991 instances of normal traffic.

Although training the model LSTM model took significantly more time than that of the LR model, it was 9% more accurate. After that, another work [259] used the techniques from the Aegean Wi-Fi Intrusion Dataset (AWID). This dataset includes normal traffic (1.633.190 instances for training while 530.785 instances for testing), injecting attacks (65,379 instances for training while 16,682 instances for testing), impersonation attacks (48,522 instances for training while 20,079 instances for testing), and flooding attacks (94848 instances for training while 8097 instances for testing). For multi-class classification, LSTM outperformed softmax with a 14% increase in accuracy. Similar research conducted by Abeshu et al. found that IoT devices' resource limitations rendered them vulnerable to DoS attacks [61]. In a widely dispersed network like the IoT, traditional machine learning techniques are less scalable and less accurate for detecting cyber-attacks. Data from billions of IoT devices allows deep learning models to outperform shallow algorithms in learning.

The authors of [61] claimed that the majority of the used deep learning architectures applied pre-training for feature extraction. It enabled the detection of abnormalities and therefore decreased a network administrator's workload. Nevertheless, their work concentrated on networked deep learning through model and parameter exchange for fog computing applications. Fog computing decreased the IoT devices' load on processing resources as well as storage space. As a result, it is the perfect location for detecting an intrusion. For fog-to-things-based computing, parallel computing is required for the traditional Stochastic Gradient Descent (SGD) algorithm. Consequently, a vast number of data produced by the IoT will choke the centralized SGD. Accordingly, the study provided a distributed deep learning-driven IDS based on a dataset like NSL-KDD, in which stacked auto-encoder (SAE) was applied to extract features, as well as soft-max regression (SMR), was employed for classifying the data while SAE performed better as a deep learning than existing shallow algorithms according to accuracy, FAR, and DR. There is evidence to support both the authors of [62] and [61] assertions that deep learning models outperform shallow machine learning algorithms. Tan et al. [64] attempted to identify DoS using a triangle-area-based technique in Multivariate Correlation Analysis (MCA). The data that made it to the target network was utilized to develop features that minimized overhead. Geometrical connections between two different characteristics were identified sby applying the "triangle area map" module to improve the accuracy of zero-day attacks for detection. According to the researchers of [64], they applied Earth Mover's Distance (EMD) to determine the differences from observed traffic to a pre-built ordinary profile, which they believed would help them improve their results from [64]. Using the KDDCup99 and ISCX datasets, MCA was applied to extract characteristics from network traffic and assess the findings for anomalies. Their findings are based on a sample-wise correlation of 99.95% on KDD data as well as 90.12% on ISCX. In any event, neither the size of data in the research nor its effect on various sample sizes was revealed. MCA was not a practical technique since the change expected was not linear. In the Internet of Things, a botnet attack is a different type of DoS attack in IoT. The authors of [70] created an IDS that integrates ANN, DT, and NB to fight botnet attacks against DNS, Message Queuing Telemetry Transport (MQTT), and HTTP. It was chosen to utilize ANN, DT, and NB to better differentiate between malicious and benign vectors since their cross-entropy values were similar. The detection rate and false-positive rate were used as performance indicators, and their proposed ensemble beat each individual algorithm inside it. The accuracy was 99.54 percent on the UNSW dataset and 98.29 percent on the NIMS dataset. MiTM attacks, which are very similar to DoS attacks, are one of the most frequently occurring attacks on the network in IoT.

Numerous technological solutions have been suggested for different application scenarios in connection to this. Due to the fact that traditional feedforward neural networks are incapable of capturing time-series and sequence data owing to their causal structure, an impersonation attack on smart healthcare was prevented with the use of an LSTM-RNN, according to the researchers [79]. Additionally, the researchers were able to address the vanishing gradient issue associated with the RNN method and improve accuracy. The predicted value was first calculated using a three-month log of the dataset (for a diabetic patient who is receiving insulin injections). DL and gesture recognition were merged if the estimated and expected dosages varied by more than a certain threshold. They lacked, however, a detailed understanding of the model and analysis. The researchers of [81] employed Physical Unclonable Function (PUF), a unique silicon chip feature that may be applied as a foundation for radio frequency (RF) communication authentication to defend against impersonation attacks. Moreover, the authors were able to identify the device and train their system on it as a consequence of these offsets, all before knowing the device's degree of accuracy. The evaluation metrics were analyzed using the ANN MATLAB toolbox. According to the simulation results, machine learning can help detect 4,800 transmitter nodes with 99.9 percent accuracy and 10,000 nodes with 99.9 percent accuracy under different channel conditions. As indicated, multi-factor authentication may be used alone or in combination with other security measures. In a secure server, PUF's intrinsic and low-cost nature enables the storage of the physical values of each wireless sensor rather than utilizing existing key-based authentication. Nonetheless, the researchers made the incorrect assumption in their approach that data saved on the PUF server is safe. Aminanto et al. [80] extracted features by applying C4.5, ANN, and SVM, with ANN acting as a classifier [80]. The technique of deep feature selection and extraction started with the extraction of features using SAE, followed by feature selection applying C4.5, ANN, and SVM, and finally by classification using ANN. The accuracy of the research was 99.92 percent due to the use of the AWID dataset, which had the lowest accuracy for impersonation attacks in a prior study, accompanied by the work of [259].

According to Statista [260], the global mobile phone user base will surpass three billion by 2020. Due to the increased usage of mobile phones, they become increasingly vulnerable to virus attacks. According to the work of [72], OpCodes may be employed to differentiate between safe and malicious software. As a consequence, global feature selection introduces inefficiencies and may potentially reduce system proficiency, especially for the imbalance dataset. Information Acquired by Class (CIG). They claim that no one has ever tried to combine OpCode with DL for IoT previously. Deep convolutional networks and Eigenspace techniques were used, and the accuracy was 99.68 percent, while the recall and precision were 98.37% and 98.59%, respectively. Similarly, Wei et al. [75] utilized dynamic analysis to extract malware features. They trained the classifier using functional application classification on clean and damaging data and then used kNN to divide the data into recognized categories during the testing phase. We performed tenfold cross-validation using the J48 decision tree and NB. Depending on the performance metric used, this study achieved 90 percent accuracy. The work of [73] applied static analysis methods rather than dynamic analysis (see [75]) to extract features, taking into account all APIs that had not previously been studied. The most commonly used qualities by earlier researchers served as a roadmap for developing new characteristics. They offered 98.9% accuracy using the dataset on the second-largest malware testbed. With the advancement of sophisticated infiltration techniques, static analysis became outdated, necessitating the use of a dynamic methodology [74]. The attackers utilized static analysis because they exploited deformation technologies to avoid recognition, while dynamic investigation approaches showed promise due to their resilience to similar tactics. The creators of [74] created the EnDroid framework in response to these issues. When it came to categorizing the data, the suggested model applied "Chi-Square", feature extraction and a combination of five ML algorithms (linear SVM, decision tree, boosted trees, random forest, and extremely random trees), with LR serving as the meta-classifier. The dataset was created by combining the Drebin and AbdroZoo databases, yielding a 98.2 percent accuracy.

For example, Wang et al. claimed that static string characteristics such as API and permissions use retrieved from applications were the basis for the majority of current malware detection literature [261]. However, due to the increasing sophistication of malware, relying only on a static characteristic may lead to a false positive. To identify Android malware, the DriodEnsemble model used a combination of string and structural characteristics. RF, KNN, SVM were applied to test the model against 1,386 good applications and 1,296 bad apps. However, using just string characteristics, the research was able to achieve 98.4% accuracy, which was higher than the recognition accuracy of 95.8% achieved by applying only structural features. It's a general method that looks for anything out of the ordinary and flags it as a potential security risk. Many researchers [262-264] have tried to use machine learning techniques to create safe intrusion detection systems (IDS). To help with this, Javaid et al. [263] used an unsupervised DL method known as STL, which relied on SAE and SMR as its foundation. Two-class classification outperformed SMR using the NSL-KDD dataset; it was superior to five-class classification by a wide margin. It was suggested by Ambusaidi et al. [262] to use Mutual Information (MI) in an ML-based multi-class classification. Linear Correlation Coefficient (LLC) was utilized for the linearly dependent variable in Mutual Information Feature Selection (MIFS). The authors utilized FMIS+MI for the non-linear dependent variable, modifying the pre-existing MIFS method [265] and demonstrating their originality. An additional motivation for doing this research was that prior studies had failed to explain the processes. Kyoto 2006+, NSL-KDD, and KDDCUP99datasets were used to compare performance, while accuracy, F-measure, FPR, and DR were used as metrics. Anomaly detection using LSTM was the focus of Maimo et al. [264]. DBN and SAE models (where the prediction may be calculated by utilizing matrix operations after an activation function) were used to reduce features because of their comparable structure, while features were extracted from flows of networks employing weighted loss functions [264]. The authors claim to have achieved up to 95% precision after utilizing the CTU-13 botnet dataset to build their model [264]. ML algorithms have been used in research that claims to decrease cyber-attacks successfully. In contrast, a past study [266] used deep feature embedding learning (DFEL) as it was faster than conventional machine learning algorithms for training data. Their approach was compared by utilizing the datasets from UNSW-NB15 and NSL-KDD, and the recall value of the Gaussian Naïve Bayes classifier improved between 80.74% and 98.79%, while SVM's runtime was decreased substantially from 67.26 seconds to 6.3 seconds as a result. The previous IoT security methods were also centralized and cloud-based, which resulted in significant high power consumption and latency for end devices [267]. Fog computing was utilized in two stages to build the suggested IDS for IoT in a distributed manner. The identified threats were then compiled and evaluated on a cloud server in the second phase. The novel method outperformed the current NB, ANN, and conventional ELM in terms of accuracy, FRP, and TPR. Fog computing-based attack detection was shown to be quicker than cloud computing-based attack detection in the experiments conducted on the Azure cloud. However, no current ML/DL-based fog-computing algorithms were utilized to compare the findings of the research.

Table 1: Summary of several security solutions in IoT using DL and ML

| IoT Application | Threat | Dataset | Kind of threat | Using Algorithms | Accuracy | Reference |
|---|---|---|---|---|---|---|
| Healthcare | MiTM | Private | Impersonation | LSTM RNN | - | [79] |
| IoBT | Malware | Private | Code Injection | DCN | 98.37% | [72] |
| WiFi | MiTM | AWI | Impersonation | ANN | 99.92% | [80] |
| Fog | DoS | NSL-KDD | Flooding | Softmax | 99.20% | [61] |
| NIDS | Anomaly | Kyoto 2006+ | Anamoly | Softmax | 88.39% | [263] |
| Android | Malware | Drebin and AbdroZoo | Malware | ensemble+LR | 98.10% | [74] |
| Fog | DoS | AWID and ISCX2012 | Flooding | LSTM | AWID (98.22%) and ISCX2012 (99.91%) | [62] |
| IoT | Botnet | NIMS botnet and UNSW-NB15 | Flooding | Adaboost | UNSW-NB15 (99.54%) | [70] |
| Android | Malware | Private | Malware | kNN, C4.5, NB | - | [75] |
| RF Communication | MiTM | Private | Impersonation | ANN | 99.90% | [81] |
| Android | Malware | Multi-sources | Malware | ensemble | 98.40% | [261] |

### 7.2. Privacy Solutions

Table 2 shows some suggested privacy-preserving ML and DL methods. A MiTM attack compromises both security and privacy. Several number of works utilize ML techniques to defend against various MiTM threats. Table 2 shows some suggested privacy-preserving ML and DL methods. A PHYlayer authentication scheme based on IAG lowered the overall communication burden and improved detection precision. When working with an updated dataset, the researchers were able to improve FAR, DR, and computing costs. There was also a problem with the wearable device that was highlighted by Aksu et al. [268] in addition to user authentication difficulties. However, it is necessary to authenticate the device itself. Similar to MiTM devices, these devices may be used to authenticate users. However, if anything goes wrong in the background, it may end up giving the attacker complete access to the system. A more powerful base device can be reached only via Bluetooth with encryption and authentication. It was considerably more secure to utilize hardware-based fingerprinting since the encryption and device name secrets might be stolen so readily. The suggested framework in [268] made use of a timing technique of classic protocol packet-based and inter-packet timing-based analysis in Bluetooth. This process has a structure of four stages. Bluetooth classic packets were first captured. The characteristics were then retrieved in a second phase. The fingerprints were produced in a third stage by using probability distributions. The saved fingerprints from step three were also matched to any fresh incoming data from wearable devices as the last step in order to identify any unfamiliar wearable devices. The study claims to have achieved 98.5 percent accuracy by selecting the best algorithm from a set of twenty training results. You'll need a large amount of data to build an ML model. E.g., we can utilize past patient data to predict outcomes for each new patient. Patients, on the other hand, are apprehensive about disclosing their personal information. According to [269-271], research has attempted to address these problems. Non-linear kernel SVM was used in [269] to effectively categorize medical data while maintaining the privacy of both the service provider and the user data model. Zhu et al. [269] said that they were able to obtain 94% classification accuracy using their system, apart from sacrificing privacy. Users' private information and model outputs were categorized as model-privacy problems and learning-privacy problems, respectively, by researchers in [270]. This study relies on gradient values instead of actual data or assumes that the learning model is private, but the learned model is public or uses complex encryption techniques. Previous research has depended on these approaches [270]. The authors of [270] presented a uniform Oblivious Evaluation of Multivariate Polynomial algorithm that lacked complex encryption methods in contrast to the other research. In the end, their findings showed that the categorization data and models they learned were safe against a variety of intrusions. Model privacy was the subject of investigation [270]. However, the issue of student privacy was not addressed. This problem was addressed by Ma et al. [271] said that although utilizing the public key to encrypt any user data was a popular technique for maintaining privacy, it came at the cost of key management. In the cloud, a cloud service provider delivers encrypted client data to a data training system that doesn't know what's being trained on. This is their suggested approach. They concluded from their analysis of the privacy-preserving DL Multiple-keys (PDLM) that it had less efficiency than traditional non-private methods while still preserving privacy. To classify data privately, they used hyperplane decision-based private methods like decision trees and Naive Bayes, together with private Naive Bayes and decision trees. In a related study, it was discovered that the number of user-server iterations could be cut in half without compromising privacy. People's lives have been enhanced by Facebook and Twitter, yet privacy concerns have arisen as a result. Blacklisting methods were used by a number of businesses to screen out malicious traffic. According to this study, 90 percent of people will be victims of these attacks even before they are prohibited. Machine learning algorithms were used to evade these attacks successfully. However, because of their slower pace of learning, these algorithms were inefficient in real-time. The authors of [272] described a multistage detection framework employing deep learning, in which the results were first detected at a mobile terminal and subsequently sent to a cloud server for additional computation. The authors stated that by utilizing the Sino Weibo dataset and CNN as a categorization technique, they obtained an accuracy of approximately 91 percent. When looking for a solution, researchers used distributed ML methods and collaborative IDS, as well as ideas of dynamic differential privacy to protect a training dataset.

**Table 2:** Summary of several security solutions in IoT using DL and ML

| IoT Application | Threat | Dataset | Kind of attack | Using algorithms | Accuracy | Reference |
|---|---|---|---|---|---|---|
| Wearable devices | MiTM | Private | Authentication | best of 20 | Precision: 98.5% | [268] |
| MSN | Anomaly | Sino Weibo | Spam | CNN | 91.34% | [272] |
| Distributed Systems | Data Privacy | Real world | Multiple | OMPE | - | [270] |
| Cloud | Data Privacy | - | Data Leakage | SGD | 95% | [271] |
| WSN | MiTM | Private | Spoof detection | DQ, QL | - | [273] |
| Healthcare | Data Privacy | Real world | Multiple | SVM | 94% | [269] |
| MiTMO Landmark | MiTM | Private | Spoof detection | Softmax | - | [55] |
| VANET | Data Privacy | NSL-KDD | Inference attack | LR | - | [274] |

## 8. New insights in Machine and Deep Learning for IoT Security

Entrepreneurial or commercial off-the-shelf IoT devices are usually supported with the solutions of software that are insufficient to protect every IoT device or system [275, 276]. Since the IoT has many different use cases, the software-level security is poor. IoT security is an issue that some researchers [21, 277] are concerned about regarding privacy and security.

### 8.1. Data Privacy

IoT security is challenged by data privacy because of the significant risk of vulnerability, according to most research [278, 279]. Unauthorized access to data, Eavesdropping, data fabrication, data alteration, and unlawful remote access using devices are some of the vulnerabilities [280]. As an example, personal information, such as names, addresses, phone numbers, insurance policy numbers, and bank names, is always at risk when it is stored on the cloud. Many IoT devices and apps, on the other hand, provide access to important information that might be used by attackers to gain access to the system. As a result, sensitive personal information that is unprotected and unencrypted may be exposed to an unauthorized party.

### 8.2. Vulnerabilities in IoT

IoT devices are now prone to several vulnerabilities. Services and data in the IoT may be susceptible to attack because of their sensitive nature [281]. Many IoT systems and a highly complex ecosystem in IoT can have increased risks from significant problems in cloud security [282]. Centralized management platforms and older systems pose substantial security risks for IoT devices [283]. When it comes to application layer security, it's possible for users to create weaknesses. One or more of many types of defects exist, including inefficient input/output filtering, poor encryption, and tampered authentication mechanisms. Few examples of vulnerabilities in IoT security are as follows.

(i) Weak, guessable, or hardcoded passwords: To get access to a system, a user must utilize credentials that are readily brute-forced, publicly accessible, or impossible to modify [284-286]. Credentials that are both hard-coded and integrated into IoT devices constitute a threat to both IT systems and the IoT itself. Hard-coded or guessable credentials are also a benefit to hackers who want to target the device. In addition, the malicious attacker can already have access to the password of a machine if it has default passwords. In order to prevent unauthorized access, devices connected to the IoT should have measures in place, i.e., password expiry, password difficulty, and one-time password account lockout that compel users to alter the default credentials. The producers of IoT devices should, as a consequence, provide them with strong passwords straight out of the box to prevent security flaws.

(ii) Inadequate protection in privacy: Insecure, inappropriate, or unauthorized use of the personal information of users kept on the ecosystem in a device may lead to IoT security flaws [287-290]. Since IoT devices might be vulnerable, proper privacy protection must be provided for them.

(iii) Vulnerable interfaces in the ecosystem: Backend application programming interface (API), web, mobile or cloud interfaces outside of the device ecosystem may be exploited to get access to the device or its components, making IoTs vulnerable to attack [291-294]. Lack of authorization and authentication may also lead to IoT vulnerabilities [295, 296], as can inadequate encryption or lack of encryption [297], as well as a lack of output and input filtering [298]. It is possible to protect a connected device as well as create data via custom device authentication. A digital entity (computer, IoT device, etc.) may also securely send data to authorized recipients using digital certificates.

(iv) Absence of any kind of hardening measures: As a result of the absence of hardening measures, attackers may get critical information that might be used to help in remote attacks and achieve local control in IoT-based systems [284, 287, 299]. Account lockout, password, and complexity that forces anybody setting up a device to modify the default credentials are among the physical hardening methods [300-303]. Because of this, physical precautions such as security paradigms are needed to guard against IoT attacks and vulnerabilities.

### 8.3. Authorization, authentication, and identification

IoT devices have a number of security issues, including the inability to be identified, verified, and granted access to the network. The authorization, authentication, and identification of IoT devices is a major concern for many researchers [6, 304, 305]. Many IoT devices do not allow a single device to be uniquely identified, authenticated, and authorized, which makes things incredibly complicated.

Furthermore, there is a difficulty with authentication. To prevent unauthorized users from having full access to a network's resources, some kind of access control is required. It was a survey of IoT communication protocols conducted by the authors of [306]. It brought to light the fact that there are now just a few protocols that guarantee users' safety and confidentiality. As a result, additional research is required to improve a framework that can provide IoT device users with privacy and security.

### 8.4. Behavior-based mobile device authentication

Behavioral authentication on mobile systems identifies an individual according to unique qualities, including biometric authentication, that utilizes patterns exhibited while networking with a system including a computer, tablet, or smartphone that contains a keyboard as well as a mouse. A secure authentication system is necessary to restrict access to tablets, cellphones, e-readers, smart watches, as well as laptop computers. Laptops, desktops, mobile phones, and tablets are no longer just tools for people; they are increasingly taking on their roles. These technologies have unlocked up different ways to interact, play, and work. Because of their small size, they're easy to carry about in pockets, handbags, or other bags. However, mobile devices are susceptible to a variety of issues. The security and privacy of the user is at risk if the device is lost or stolen. It is possible to get threats from both strangers and close friends. Similarly, mobile devices are readily lost because of their mobility and portability. Users' private life and personal information might be made public if a thief gains access to these devices. They may also be vulnerable to extortion or blackmail. Moreover, a biometric technique aims to identify and detect the user. The United States National Science and Technology Council's Subcommittee on Biometrics separates biometrics into physiological categories and behavioral [307, 308]. As the name suggests, behavioral biometrics is concerned with identifying and quantifying human behavior patterns. The identification approach based on physiological features is quite accurate. Physiological biometrics, on the other hand, focuses on physical characteristics of the human body, such as a retinal or fingerprint scan. Conversely, behavioral biometrics denotes behavioral aspects of the human body. Behavioral biometrics analyses data, including a user's screen pressure, navigational patterns, mobile or mouse motions, gyroscope position, typing speed, etc. Behavioral biometrics recognizes a subject by employing behavioral qualities. Each subject is projected to differ from all others when investigated using one or more of these characteristics. Additional human aspects and behavioral biometrics attributes and verification methods include gait analysis, keystroke dynamics, touchscreen, voice ID, hand waving, mouse usage characteristics, signature analysis, cognitive biometrics, electroencephalogram (EEG), profiling, and an electrocardiogram (ECG). An important benefit of behavioral biometrics is that it may be used to authenticate users without the requirement for additional hardware [307]. To put it another way, adopting behavioral biometrics rather than physiological biometrics is more cost-effective. Analyzing an individual's physical characteristics is possible via the use of a variety of biometric tools, including retinal or iris scans, face identification software, and fingerprints. In the same way, it involves assessing how a person uses their pen, as well as their personality characteristics and other aspects of their everyday conduct. Authentication and identification are two of the most common uses of biometric technology. More secure systems might be created by using various authentication methods.

Pin/password, authentication using pattern, speech recognition, face recognition, iris-based authentication, and fingerprint recognition are among the authentication systems mentioned as follows;

(1) Fingerprint recognition: Fingerprint identification may also be accomplished with the use of a secret sign. It is described as a precise pattern of finger movement over the screen. Users may authenticate themselves using this pattern as a kind of biometric authentication. A biometric is a trait of a person's physical or mental makeup that cannot be duplicated. Using biometrics, it is possible to tell one individual from another. In other words, it's a way of figuring out someone's identity.

(2) Iris: The colorful part of the eye around the pupil is called the iris. It is biometrics that is often regarded as reliable. Because each person's iris has its unique patterning, a blood test may be performed accurately, quickly, and simply. The iris may be matched using a picture since the eye is a visible organ [309]. Many airports in the UK, including Manchester, Heathrow, Birmingham, and Gatwick, began using iris scanners in 2004 as part of a nationwide rollout. Their usage was later phased out since it was believed to take more time than ordinary passport inspections to complete the process. According to firms like EyeLock, the IoT and autonomous automobiles will benefit from iris scanning. Each individual can only have two different iris pictures since it is a fixed characteristic. Iris scanning equipment may take up a lot of room. Close closeness is also required.

(3) Pin/Password: A personal index number (PIN) or a secret pattern is the current form of authentication for cellphones, tablets, and laptop computers. Typically, a PIN requires four or more numbers to be entered by the user for verification. This code must be entered correctly for the user to access their device.

(4) Fingerprint: Fingerprint scanning is one of the least expensive biometrics, making it an attractive option for many organizations. Fingerprint images may be captured using a tiny camera that can be incorporated into mobile devices like wearables or smartphones, making it very convenient. This means that mobile apps on devices with this hardware may be authenticated using this way. Because mobile devices have limited typing skills, password authentication is often unpleasant. If the child's fingerprints change, this is not the best biometric.

(5) Facial recognition: In comparison to other technologies, facial recognition is a non-intrusive and low-cost option. Due to the widespread use of selfies, the smartphone is well-suited for face recognition. As a result, smartphone makers have made significant investments in front-facing cameras. Using the device's screen, people may check to verify whether the camera is taking a picture of the right region of their faces. Some UK airports are already using facial recognition technology at ePassport gates. MasterCard's self-service payment app also takes advantage of it. A possible drawback is that illumination changes might alter the picture. When a person grows older or trims their hair, their facial features alter as well. Plastic surgery on the face has a significant chance of altering it. A significant difficulty for automated face identification is the failure of several face recognition algorithms to distinguish faces after cosmetic surgery. Additionally, attackers may exploit facial recognition technology [310, 311].

## 9. Challenges, Limitations, and Future Directions

Machine and deep learning algorithms have only recently been developed and are not intended for use in cryptographic applications. Two previous studies [130, 144] show, for example, that ML can be applied to hack a sample attack using SVMs and cryptographic constructs. Similarly, developers of [312] taught DL algorithms to decode cryptographic frameworks and concluded that DL would do so. Machine learning (SVM and RF) and logical process profiling algorithms were outperformed by CNN and AE algorithms. RNNs have previously been shown to be capable of learning decryption. The study of successful internal representations of this cipher may also be used to decode the enigma machine on an RNN with a three-thousand unit LSTM; The results also suggest that deep learning algorithms such as RNN can detect and manage polyalphabetic cipher algorithms for cryptanalysis [313]. Machine learning/deep learning research has the potential to advance the advancement of the Internet of Things.

As an enormous number of intelligent items are linked to IoT devices, it is critical that the endpoints of such devices be secure. Profiles, explicit trust connection, timestamping protocol, privileges, encoding, and so on all need robust authentication protocols [19, 314, 315].

### 9.1. Limitations of ML in IoT

The one disadvantage of simple ML approaches is that they need a large amount of data for model testing. The studied model is then used to approximate or categorize real-world implementation outcomes. However, it should be remembered that the whole procedure does not capture the whole spectrum of data characteristics and facilities. In this case, DL methods were used to address the shortcomings of machine learning strategies. Since DL can process vast amounts of data and its algorithms are flexible when the volume of data increases, model testing is advantageous and predictive accuracy can be enhanced. High-level functions and contrasts are derived dynamically and hierarchically from input data by DL. The majority of IoT implementations produce blank or half-marked results. Unlabeled data may be used by DL in an unsupervised manner to reveal valuable trends. Typical machine learning algorithms are only successful where there is a large amount of data on the label [46].

### 9.2. Limitations of DL in IoT

We conducted a thorough review, which revealed that existing research needs to be changed in order to reach higher protection requirements in IoT settings. Security issues are essential because authorization, entry security, system security, data integrity, intrusion detection techniques, and packet extraction both play a role in the detection of anomalies. Security concerns are severe. The specific IDS algorithm, data preprocessing, function extraction, and the optimal set of features are all important factors in DL-based anomaly detection. Flexibility and planning are also common issues for profound learning approaches. The authors of [316] examined various DNN models and discovered that small precision improvements take a long time. Moreover, tuning the parameter is a significant issue since the number of layers and accuracy are linearly related. Many hyperparameters are thus expected if deep learning methods that are highly sensitive to data structure and size are to start optimally. The research challenges in the environments of DL-based IoT security contain:

(i) End-to-end safety (integrity, access management, authentication, confidentiality, and intrusion detection systems);
(ii) Data pre-processing, optimum function selection, and extraction.

### 9.3. Challenges of ML

As addressed further below, an obstacle is the insufficient collection of data for data-driven ML and DL methods.

**(i) A scarcity of testing datasets:** Data sets are available for the effective use of DL and computer education solutions. To validate and evaluate the output of various profound learning and enhancement learning algorithms, authentic databases from the actual physical world are used. The data includes sensitive and personal knowledge that not only differentiates individuals but also their habits and way of life. Data created by BAN and other healthcare apps, for example, can jeopardize consumer safety, while data from intelligent homes can influence lifestyle and behavior. As a result, it is important not to jeopardize consumer safety when using ML and DL. Numerous methods of anonymization were used to anonymize data until it was used for analytics; nevertheless, the study revealed that these techniques could be hacked and models abused by adding fake data. It may be difficult to gather data while maintaining secrecy and privacy. Furthermore, issues such as how machine learning and profound learning algorithms would be applied, as well as the extent to which machine learning and profound learning algorithms may protect privacy, must be addressed. As a result, it is critical to investigate machine learning and IoT network deep-learning analytics strategies for data security and consumer privacy safety. Please keep in mind that simulation data cannot accurately represent real IoT scenarios in the universe. Furthermore, generating synthesis data to train and test deep learning models may be computationally costly.

**(ii) Data imbalance:** When attacks are uncommon in an IoT environment, the data sets obtained for machine learning or profound learning are more likely to be unbalanced. The dependability of attack classifiers and intrusion detection systems would have a significant effect on these various data sets.

**(iii) Data convergence:** It would be necessary to combine data from various IoT devices and network modules in order to construct machine learning and profound learning models. This may be daunting since data from various sources can differ in modality and granularity, complexity and falsity.

### 9.4. Challenges of DL

ML, which is a technique for extracting information from results, has been used for both malicious and benign purposes. It has been discovered that future adversaries allow effective use of these ML and DL-based learning algorithms to crack cryptographic secrets. For instance, the authors employ recurrent neural networks for cryptanalysis. Additionally, erroneous data inputs to the ML algorithm result in an inefficient operation of the whole learning-based framework. Oversampling, an insufficient testing dataset, and function extraction are all issues to consider when applying knowledge to smart ecosystems.

### 9.5. Future Directions of ML

Artificial intelligence and machine learning have provided a major contribution to computer security progress. They also improve the quality of our daily lives by using, inter alia, IoT devices, intelligent homes, and smart automobiles. An advanced protection framework cannot be deemed complete before AI, and ML components are used. AI and ML solutions can mainly help identify similarities between specific previous attacks and include an automatic warning when any similar danger is identified. The most valuable feature of AI/ML is that it can consistently discern user behavior, changing use patterns, and many other anomalies [3, 317]. One of our testing recommendations, which security experts have agreed to, is to standardize the data packages accessible in order to facilitate the decoding and interpretation of data through machine learning solutions. Our data collection is calculated in exabytes. Upon specifying and optimizing data sets, machine learning algorithms can be beneficial in the protection against cyber threats. We recommend that a fine line be drawn between agreeing on a supervised solution based on features derived from our data collection on the basis of our proposed research solution. While AI and machine learning systems should run independently without human intervention, a small amount of human input should be provided to maintain the system balanced and functional. Although it is limited to creating a hybrid detection model for combating and mitigating IoT cyberattacks in a host and network infrastructure environment, we also suggest using different algorithms such as Eclat and Apriori to warn users of cyberattacks.

### 9.6. Future directions of DL

Building modern IoT network architectures with protection protocols including authentication, access control, confidentiality, and intravenous system detection is a successful solution for end-to-end safety. New IoT architectures must prioritize quality of service over efficiency, and they must incorporate evolving paradigms such as SDN and fog-enabled IoT. Optimization algorithms such as genetic algorithms (GA), bacterial foraging optimization (BFO), particle swarm optimization (PSO), and attribute extraction and selection techniques, as well as parameter tuning, can be used. Hybrid deep learning techniques may be used to increase performance without significantly raising computing time. Blockchain technology, which uses deep learning, may also be used to improve IoT stability. Blockchain technology is a relatively new solution to ensuring the secrecy and security of distributed records.

### 10. Suggested IoT Security Practices

While past IoT security concerns have been addressed, there are more considerations that need to be made, such as the following suggestions.

- A recognized IoT cybersecurity framework based on industry experience, standards, and proper procedures provided by regulatory bodies should be used.
- Second, IoT devices should not rely only on the network firewall to prevent malicious communication.
- Generate a Cybersecurity/IoT incident response strategy and immediately assign the router a name.
- Weakness examinations of devices that are linked to remote systems are very important.
- It is good to periodically update the default login credentials and double-check all connected devices.
- IoT systems must be partitioned or isolated to decrease the number of points of attack.
- Threat intelligence must be monitored and shared. In addition, it is critical to scan all software to

- ensure that the network does not have any security holes.
- In order to digitally fence networks and devices, it is essential that security software be installed and objects and containers are added.
- People, businesses, and governments need to keep an eye on and exchange information about threats.
- Other attack detection measures, such as DDoS, IP spoofing, and so on, may be implemented.
- Devices and networks must be updated and patched on a regular basis.
- Avoid adding devices to the network that use default passwords or have known security flaws.
- Device apps and controllers need to have their access credentials verified.
- Biometrics and robust validation should be utilized for access control.
- When linked to a system, use machine validation and IoT messaging encryption, especially for data in transit.
- In order to protect the LAN from the Internet, firewalls already in use need to be upgraded to more powerful models.
- If you're using Wi-Fi, be sure you're using a secure router and using passwords that are strong and unique. Wi-Fi security also necessitates the use of high-quality encryption.
- Make use of a variety of security measures, including antivirus.
- Whenever feasible, make a copy of all of your data. Inbound connections to connected devices should be disabled by default.
- All data should be safeguarded from unwanted access, both while in transit and while it is stored.
- Devices must be able to delete or reject data storage items with ease. Secure USB ports, for example, should never be used by systems with exposed external interfaces.
- It is possible to hire security specialists or to hire cloud security professionals.
- Predictive analytics, real-time monitoring, and auditing in the long term
- To ensure the safety of all employees and the general public, security awareness training and public exposure to hacker and intruder techniques and tactics are critical.
- The IoT should be more restricted to intruders rather than more lenient. Only trustworthy endpoints should be used to connect with IoT items, according to industry standards.
- There should be a clear emphasis on eliminating security concerns, such as illegal hacking or operation, environmental risks, tampering, and system malfunctions in IoT systems.
- Limiting the consequences of a security breach on a possible attacker, such as allowing personal identifying information and assuring rapid discovery and prompt handling of any breaches.

The objective of securing the IoT has yet to be fulfilled, despite different initiatives. IoT security is still a difficult topic to solve. However, the use of cutting-edge artificial intelligence-based cybersecurity systems may considerably deter invaders.

## 11. Conclusions

Traditional security and privacy strategies have many problems linked to the complexity of IoT networks. DL and ML technology can be used to adjust IoT devices to our real life. The review considered several types of IoT threats. DL and ML are addressed several potential solutions for ensuring IoT security. A number of DL and ML models are illustrated with their application in IoT security. This review discusses the state-of-the-art solutions for IoT privacy and security utilizing deep learning and machine learning techniques and their integration. While studying machine learning privacy and security issues, we also made an effort to develop a review of IoT threats using previous studies on DL and ML. New issues and insights of ML and DL in IoT security are addressed. Moreover, future direction, security challenges, limitations, and suggestions are included for empowering future technology.

**Funding:** This research work did not receive funding.

**Availability of data and material:** Not applicable

**Compliance with ethical standards:** N/A

**Conflicts of interest/Competing interests:** All the authors in the paper have no conflict of interest.

**Ethical approval:** This article does not contain any studies with human participants or animals performed by any of the authors.

**Consent to participate:** N/A

**Consent for publication:** N/A